\documentclass[aps,prd,reprint,secnumarabic]{revtex4-1}
\usepackage{blindtext}
\usepackage{hyperref}
\usepackage{hypernat}

\usepackage{enumerate}
\usepackage{amsmath,amssymb,epsfig,cancel}
\usepackage{mathtools}
\usepackage{color}
\usepackage{subfig}
\usepackage{graphicx}
\usepackage{caption}
\usepackage{braket}
\usepackage[english]{babel}

\usepackage{physics}
\usepackage{cancel}
\usepackage{slashed}

\makeatletter
\renewcommand\@makecaption[2]{%
  \par
  \vskip\abovecaptionskip
  \begingroup
   \small\rmfamily
    \begingroup
     \samepage
     \flushing
     \let\footnote\@footnotemark@gobble
     \@make@capt@title{#1}{#2}\par
    \endgroup
  \endgroup
  \vskip\belowcaptionskip
}
\makeatother

\begin{document}
\title{Nested Holography}
\author{Kostas Filippas}
\email{kfilippas21@gmail.com}
%\author{else}
%\author{Someone Else}
\affiliation{Institute of Nuclear and Particle Physics, NCSR Demokritos, GR-15310 Athens, Greece}

\begin{abstract}
Recently, we introduced a symmetry on the structure of angular momentum which interchanges internal and external degrees of freedom. The spin-orbit duality is a holographic map that projects a massive theory in four-dimensional flat spacetime onto the three-dimensional $\mathbb{S}^2\times\mathbb{R}$ null infinity. This cylinder has radius $R\sim1/m$ and, quantum-mechanically, its vacuum state is a fuzzy sphere. Progress shows that, first, this duality realizes the Hopf map, a fact manifest on the superparticle. Secondly, the bulk Poincar\`e group transforms into the conformal group on the cylinder. In fact, the general version of the duality yields that the dual symmetries include the BMS group, as is appropriate at null infinity. As an example, the Landau levels in $\mathbb{R}^3$ are shown to match those of a Dirac monopole on the dual $\mathbb{S}^2$, in the thermodynamic limit. This dual system is actually identified with a three-dimensional critical Ising model. The map is then realized on $N_f$ massive fermions in flat space which, indeed, are the hologram of $2N_f$ massless fermions on the cylinder. However, the dual space is really the conformal class of $\mathbb{S}^2\times\mathbb{R}$, naturally enclosing the universal cover of a conformally compactified AdS$_4$ spacetime. We argue that, in the absence of interactions, the massless fermions on the conformal boundary are in turn dual to $N_f$ massive fermions in AdS$_4$. For free fermions, all path integrals $-$the ones in $\mathbb{R}^4$ and $\mathbb{S}^2\times\mathbb{R}$ and AdS$_4-$ are shown to match. Hence, AdS/CFT duality emerges into a larger context, where one holography nests inside the other, suggesting a complete holographic bridge between fields in flat space and the AdS superstring.

\end{abstract}

\maketitle

\section{Introduction}
In \cite{Filippas:2022sna}, we introduced an automorphism of the Lorentz algebra, based on a certain kind of decomposition of bivectors, in this case of the angular momentum generator. Hence, this is a symmetry of a symmetry, a transformation that interchanges internal (spin) and external (orbit) degrees of freedom.

\subparagraph{The duality map}
The whole story is built on first principles, really. That is, we just consider Poincar\`e invariance in four-dimensional flat space, where the Lorentz subgroup through Noether's theorem implies that angular momentum breaks up into spin and orbital parts,
\begin{equation}
J^{\mu\nu}\;=\;L^{\mu\nu}+S^{\mu\nu}\;,
\end{equation}
which are complementary, sharing a common conservation law, $\dot{J}^{\mu\nu}=0$. This is a kinematic-dynamic complementarity, in the sense that $L^{\mu\nu}$ is kinematic while $S^{\mu\nu}$ is not and interacts with external fields. It can be made geometric, by means of an old-fashioned $(1+3)$-decomposition $\eta^{\mu\nu}=\frac{p^\mu p^\nu}{p^2}+n^{\mu\nu}$, where $n^{\mu\nu}p_\nu=0$, w.r.t. a massive configuration with four-momentum $p^\mu$,
\begin{equation}
J^{\mu\nu}\;=\;E^\mu\wedge p^\nu\;+\;\star( H^\mu\wedge p^\nu)\;,
\end{equation}
where
\begin{equation}
\begin{split}
E^\mu&=\frac{J^{\mu\nu}p_\nu}{p^2}=\frac{L^{\mu\nu}p_\nu}{p^2}=N^\mu\;, \hspace{1.7cm}\\[5pt]
H^\mu&=\frac{p^\nu\star J^{\mu\nu}}{p^2}=\frac{p^\nu\star S^{\mu\nu}}{p^2}=W^\mu\;.\hspace{1cm}
\end{split}
\end{equation}
This is what is called a Hodge decomposition, a generalization of the usual Helmholtz decomposition in $\mathbb{R}^3$ into curl-free and divergence-free parts. For $J^{\mu\nu}$, those parts are represented by the structure \cite{foot1}
\begin{equation}
\mathcal{S}\;:=\;\lbrace p_\nu\star L^{\mu\nu}=0\;,\;S^{\mu\nu}p_\nu=0\rbrace\;.
\end{equation}
$N^\mu$ is the spacelike coordinate \cite{footN}, $N^\mu p_\mu=0$, defining the exterior algebra $L^{\mu\nu}=N^\mu\wedge p^\nu$, while $W^\mu$ is the Pauli-Lubanski (coordinate) pseudo-vector, $W^\mu p_\mu=0$.
We call those electric and magnetic parts, respectively, since, in the rest frame,
\begin{equation}
J^{\mu\nu}\;=\;m\left(\begin{array}{cc}
0 &-N^i\\[5pt]
N^i &\epsilon^{ijk}W_k
\end{array}\right)_+\;,
\end{equation}
where $m$ is the mass of a configuration with $p^2=-m^2$. Then, holding four-momentum intact, $p^\mu\mapsto p^\mu$, we observe that the map
\begin{equation}
\begin{array}{c}
N^\mu\;\mapsto\; W^\mu\\[6pt]
W^\mu\;\mapsto\;-N^\mu
\end{array}\hspace{0.7cm}\Leftrightarrow\hspace{0.7cm}J^{\mu\nu}\mapsto\star J^{\mu\nu}\label{MAP}
\end{equation}
is a symmetry of the Hodge structure $\mathcal{S}$. This an electric-magnetic kind of duality, which we call the spin-orbit duality. This map is meaningful in three distinct ways. First, it is a symmetry of $\mathcal{S}$ and, equally, an automorphism of the Lorentz algebra, a fact illustrated in Appendix \ref{App1}. Secondly, even at this early stage it is a duality, since it preserves the conservation laws $\dot{J}^{\mu\nu}=0$ and $\dot{p}^\mu=0$, which are the equations of motion for a free relativistic point-particle. Finally, if instead of $J^{\mu\nu}$ we Hodge-decompose the U(1) field strength $F^{\mu\nu}$ bivector, this map reduces to the usual electromagnetic duality.

\subparagraph{The dual spacetime}
The map $N^\mu\mapsto\widetilde{N}^\mu:=W^\mu$ (where $\widetilde{N}^\mu$ emphasizes that $W^\mu$ is a dual coordinate) indicates that spacetime transforms, from infinity up to the point where the map becomes trivial, at $N^\mu=W^\mu$. That is, $\mathbb{R}^{1,3}$ transforms, except a world-tube of radius
\begin{equation}
R\;=\;\sqrt{W^2}\;\simeq\;\frac{1}{m}\;,\label{MollerRadius}
\end{equation}
around the configuration with four-momentum $p^\mu$. As a quantum-mechanical operator,
\begin{equation}
\hat{R}\;=\;\sqrt{\hat{W}^2}\;=\;\frac{\hbar\sqrt{s(s+1)}}{m}\;.\label{MollerRadiusQM}
\end{equation}
Incidentally (or not), this is the M\o{}ller radius \cite{Moller}: it signifies a region of noncovariance, a world-tube enclosing all possible pseudo-worldlines defined by $p^\mu$ \cite{Lorce:2018zpf}. Quantum-mechanically, it is of the order of the Compton wavelength $\lambda_C=\hbar/m$, implying pair production. Moreover, in Appendix \ref{App2} we prove that the coordinate operator,
\begin{equation}
[\hat{X}^\mu,\hat{X}^\nu]\;=\;-\frac{\hat{S}^{\mu\nu}}{\hat{p}^2}\;,\label{BulkFuzzy}
\end{equation}
an observation first made by Pryce \cite{Pryce:1948} for the relativistic center of mass and by Casalbuoni \cite{Casalbuoni:1976tz} and Brink and Schwarz \cite{Brink:1981nb} in the context of the (CBS) superparticle. This states that, for quantum mechanics of a configuration with spin and mass, space is noncommutative with fundamental scale $\lambda_C$. Summing up, given all three arguments, the spin-orbit map becomes trivial exactly where both relativistic covariance and quantum mechanics place a natural limitation in localization.\\

But what does flat space transform to? The spacelike $N^\mu=(0,N^i)_+$ has really only three spatial degrees of freedom, foliating $\mathbb{R}^3$ into hyperplanes normal to $p^\mu$. Since the dual coordinate $\widetilde{N}^2(=W^2)=R^2$, all those hyperplanes map onto an $\mathbb{S}^2$. This is also understood as the fact $\widetilde{N}^\mu=W^\mu$, whereas $W^\mu$ generates SO(3). At the same time, $p^\mu\mapsto p^\mu$ states that the timelike coordinate stays invariant. That is, spacetime transforms as
\begin{equation}
\mathbb{R}^{1,3}\;\;\mapsto\;\;\mathbb{S}^2\times\mathbb{R}\;.
\end{equation}
where $\mathbb{S}^2$ has radius $R\simeq1/m$. A more algebraic view is the following. The map becomes trivial on a certain radius, effectively acting on the domain $\mathbb{R}^3\setminus\lbrace0\rbrace$. Then, since $\mathbb{R}^3\setminus\lbrace0\rbrace$ is isomorphic to $\mathbb{S}^2\times\mathbb{R}$, the above holographic map is understood as the conformal immersion $\mathbb{R}^3\setminus\lbrace0\rbrace\rightarrow\mathbb{S}^2$. This is illustrated in Appendix \ref{App3}. Notice that $N^\mu$ transforms under SO(1,3), which is felt as the conformal group on the dual $\mathbb{S}^2$; this is a first hint that the dual space is actually an equivalence class.\\

\begin{figure}[t!]
    {{\includegraphics[width=\linewidth]{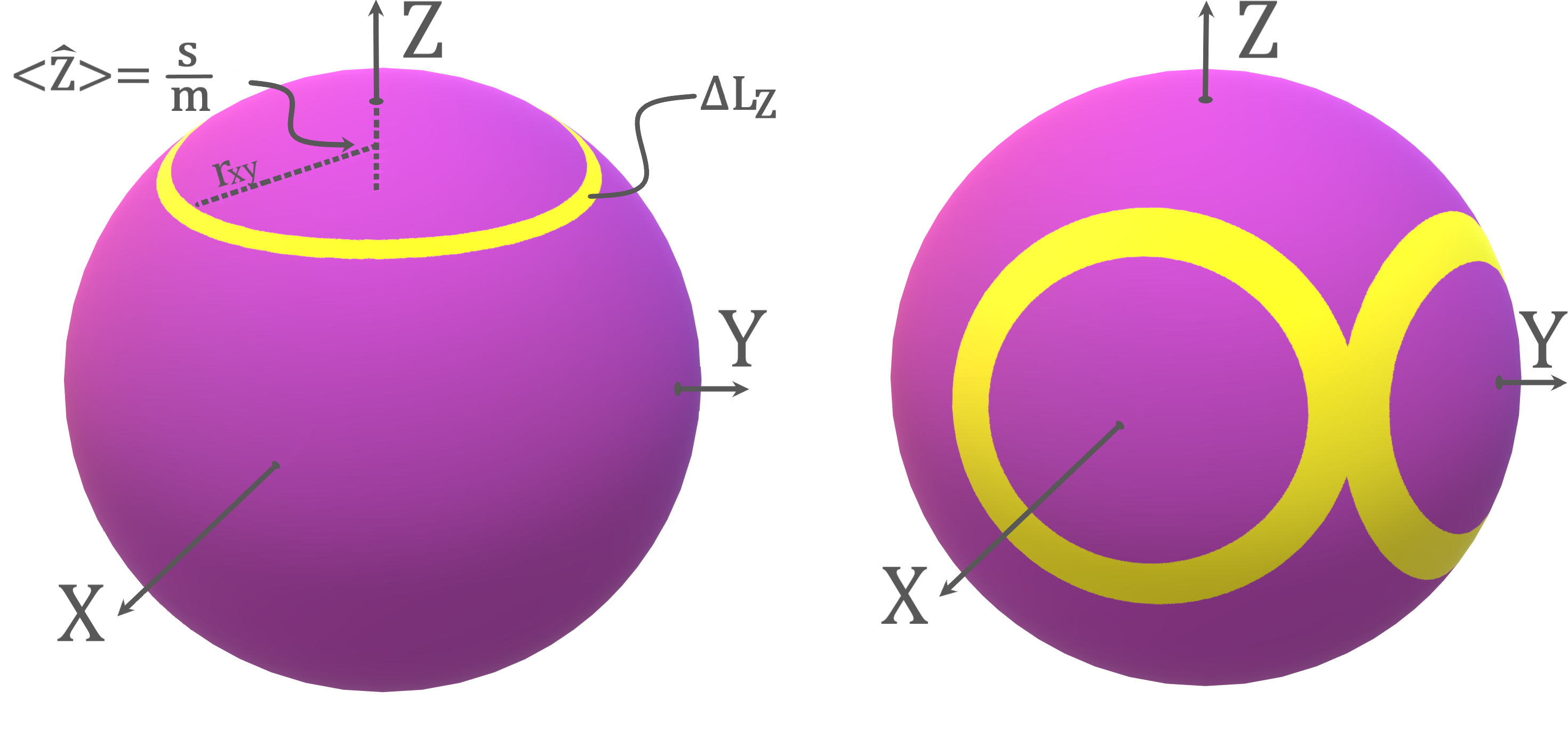} }}%
\caption{Left: a spin state $\langle\hat{S}_Z\rangle=s$ in the bulk $\mathbb{R}^3$ maps to an orbital state $\langle\hat{Z}\rangle=\langle\hat{S}_z\rangle/m=s/m$ on the dual $\mathbb{S}^2$. Accordingly, $\langle\hat{X}\rangle=\langle\hat{Y}\rangle=0$ and, thus, $\Delta Z=0$ and $\Delta X\Delta Y=r_{XY}$, which represent a $\Delta L_z$ dispersed configuration about the $z$ axis. Therefore, a maximal-spin state around an axis in bulk $\mathbb{R}^3$ is dual to \textit{dispersion ring} around the same axis on $\mathbb{S}^2$. Right: a superposition of spin states in bulk $\mathbb{R}^3$ is dual to a duet of \textit{dispersion ribbons} on $\mathbb{S}^2$. The widths of the ribbons reflect the probability amplitudes of those $\mathbb{R}^3$ spin states \cite{Filippas:2022sna}.}
 \label{RINGS}
\end{figure}

Like in $\mathbb{R}^{1,3}$, quantum mechanics on the dual $\mathbb{S}^2\times\mathbb{R}$ too is noncommutative,
\begin{equation}
\begin{split}
[\hat{X}^\mu,\hat{X}^\nu]&=\frac{i}{p^2}\left(\hat{X}^\mu p^\nu-\hat{X}^\nu p^\mu+\epsilon^{\mu\nu\rho\sigma}\hat{X}_\rho p_\sigma\right)\;,\\
[\hat{X}^\mu,\hat{p}^\nu]&=i\frac{\hat{p}^\mu\hat{p}^\nu}{p^2}\;,
\end{split}
\end{equation}
where, from now on, $\hat{X}$ refers to the dual coordinate operator \cite{footSpinNC}. Minus the third term, the first expression is a $\kappa$-deformation of the Poincar\`e-Hopf algebra \cite{Lukierski} with $\kappa=m$. Moreover, the commutator $[\hat{X}^i,\hat{p}^0]=-i\hat{p}^i/\hat{p}^0$ is known as the Newton-Wigner localization \cite{Newton:1949cq}. The important part, though, is the low-energy limit,
\begin{equation}
[\hat{X}^i,\hat{X}^j]\;=\;-i\lambda_C\epsilon^{ijk}\hat{X}_k\;,
\end{equation}
which is a fuzzy sphere \cite{Madore:1991bw} of radius $\hat{R}$. In fact, the map (\ref{MAP}) implies $\langle\hat{X}^i\rangle=\langle\hat{S}^i\rangle/m$, i.e. the dual coordinate on $\mathbb{S}^2$ is given by the spin state in the bulk $\mathbb{R}^3$. In turn, this implies that the fuzzy sphere is comprised by $2s+1$ eigenstates. We visualize all this explicitly in Figure \ref{RINGS}.\\

The interchange of internal and external degrees of freedom, $\langle\hat{X}^i\rangle=\langle\hat{S}^i\rangle/m$, yields a one-to-one correspondence between the dual Hilbert spaces. That is, the dual systems share the same density matrix $\rho$ and, thus, an identical entanglement entropy $S_{\tiny\mbox{EE}}=-\tr(\rho\log\rho)$. Surprisingly, a duality has been shown to exist for entangled systems of identical particles, under the interchange of internal and external degrees of freedom \cite{Bose}. This was proposed as a test of quantum indistinguishability \cite{Moreva}, before being generalized for distinguishable particles too \cite{Karczewski}. This so-called \textit{entanglement duality} was then applied on spin/orbital degrees of freedom \cite{Bhatti}. We suggest that this is a direct realization of the spin-orbit duality.

\subparagraph{Null infinity}
$N^\mu$ being spacelike, $-N_0^2+N_i^2>0$, it foliates Minkowski space into dS$_3$ slices. At least the part of it outside the lightcone. Inside the lightcone, the timelike coordinate $A^\mu$, $-A_0^2+A_i^2<0$, foliates space into EAdS$_3$ slices. The boundaries between those two regions are $\mathbb{S}^2_-$ and $\mathbb{S}^2_+$, at past and future infinity of the lightcone. SO(1,3) acting on the dS$_3$ and EAdS$_3$ slices is felt as the conformal group on the boundary $\mathbb{S}^2_\pm$; those spheres are the conformal boundary of the dS$_3$ and EAdS$_3$ slices. This conformal structure is a shared property of the conformal boundary and the null infinity of the lightcone.\\

\begin{figure}[t!]
    {{\includegraphics[width=\linewidth]{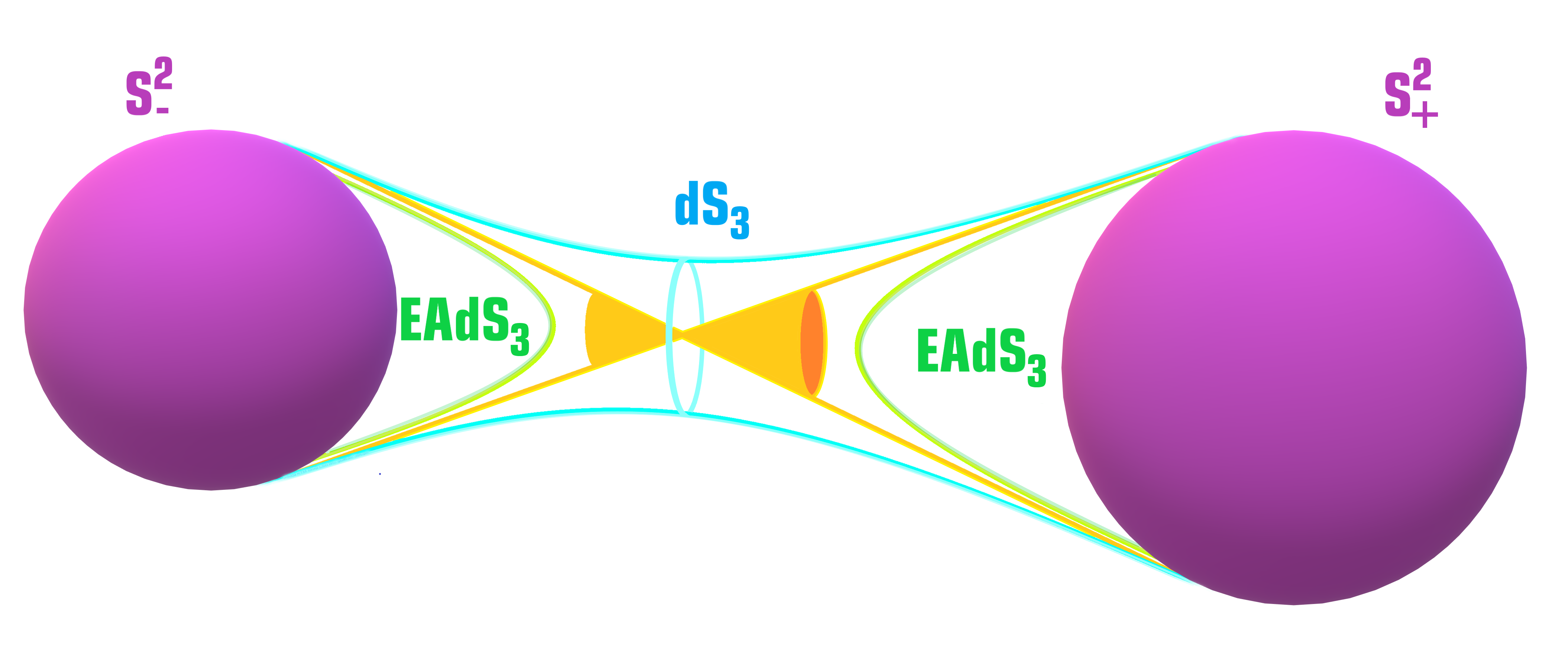} }}%
\caption{Spacelike coordinate $N^\mu$ foliates $\mathbb{R}^{1,3}$ into dS$_3$ slices, outside the lightcone. Inside the lightcone, timelike coordinate $A^\mu$ foliates space into EAdS$_3$ slices. The boundaries of those regions are $\mathbb{S}^2_\pm$, at past and future infinity of the lightcone. SO(1,3) acting on dS$_3$ and EAdS$_3$ is felt as the conformal group at null infinity, which is a conformal boundary. Spin-orbit duality is a map that projects the dS$_3$ slices on $\mathbb{S}^2_\pm$.}
 \label{SLICE}
\end{figure}

On the other hand, this view is tantamount to the structure of the spin-orbit map: $\mathbb{R}^{1,3}$ projects onto the conformal class of $\mathbb{S}^2\times\mathbb{R}$, precisely by mapping such dS$_3$ slices onto the conformal $\mathbb{S}^2$, making it natural to identify this dual cylinder with the null infinity of Minkowski space. In simple words, a massive configuration in $\mathbb{R}^{1,3}$ holographically maps onto the null infinity. We illustrate all this in Figure \ref{SLICE}.

\section{The Hopf map}
In Euclidean signature, $N^\mu$ is an SO(4) representation that foliates $\mathbb{R}^4$ into concentric $\mathbb{S}^3$'s. Then, the duality is essentially the first Hopf map,
\begin{equation}
\Big[N^\mu\mapsto W^\mu\Big]\;\;\;=\;\;\;\Big[\mathbb{S}^3\xrightarrow{\mathbb{S}^1}\mathbb{S}^2\Big]
\end{equation}
This map induces an isomorphism $\pi_3(\mathbb{S}^3)\approx\pi_3(\mathbb{S}^2)=\mathbb{Z}$ and, thus, there are infinite homotopy classes of maps from $\mathbb{S}^3$ to $\mathbb{S}^2$. Each such homotopy class has an integer attached to it, the Hopf invariant, the linking number of two 1-cycles on $\mathbb{S}^3$ which, in turn, are inverse images of different points on $\mathbb{S}^2$. Seeing $\mathbb{S}^3\cong$ SU(2) as a U(1)-bundle is a natural consequence of $\mathbb{S}^2\cong$ SU(2)/U(1).\\

The only known matrix realization of the Hopf map is the following. Consider a normalized SL$(2,\mathbb{C})$ Weyl spinor $\psi$, $\psi^\dagger\psi=const.$, defining a hypersurface $\mathbb{S}^3\subset\mathbb{C}^2$. Then, a map from $\mathbb{C}^2$ to $\mathbb{R}^3$ is
\begin{equation}
\psi\;\;\rightarrow\;\;x^i=\psi^\dagger\sigma^i\psi\;,\label{HopfMap}
\end{equation}
where $x^i\in\mathbb{R}^3$ and $\sigma^i$ are the Pauli matrices. Since $x^ix_i=(\psi^\dagger\psi)^2=const.$, this is a map $\mathbb{S}^3\rightarrow\mathbb{S}^2$. Here, $\psi$ is called a Hopf spinor and is a coordinate on $\mathbb{S}^3$. In this view, the duality map acts on spacetime as a conformal immersion $\mathbb{R}^4\setminus\lbrace0\rbrace\;\cong\;\mathbb{S}^3\times\mathbb{R}\;\rightarrow\;\mathbb{S}^2\times\mathbb{R}$.\\

This expression is useful because it manifests naturally on the CBS superparticle. In four-dimensional superspace, its action reads
\begin{equation}
S_{\tiny\mbox{CBS}}=\int\dd t\;e^{-1}(\dot{X}^\mu-i\dot{\theta}\sigma^\mu\bar{\theta}+i\theta\sigma^\mu\dot{\bar{\theta}})^2-em^2\;.
\end{equation}
Here, the duality map is
\begin{equation}
N^i\;\;\;\mapsto\;\;\;\widetilde{N}^i=W^i=\theta\sigma^i\bar{\theta}\;,
\end{equation}
which realizes the Hopf map. Together with $W^i\mapsto-N^i$, they act as just a $\mathbb{R}^3$ parity transformation in the action. Hence, this is a true symmetry for the (non-interacting) superparticle, which makes sense since $p^\mu\mapsto p^\mu$.

\section{The conformal group}
Still, we do not know the transformation rule of the bulk Poincar\`e group $G$ into its dual $\widetilde{G}$, whereas the immersion $\mathbb{R}^{1,3}\mapsto\mathbb{S}^2\times\mathbb{R}$ indicates that the dual generators must be realized in a projective manner, i.e. $G\neq\widetilde{G}$. What we do know, though, is that the map maintains the Poincar\`e conservation laws ($\dot{J}^{\mu\nu}=\dot{p}^\mu=0$), while it preserves the SO$(1,3)$ subgroup. And those facts are actually quite restrictive. The preservation of the conservation laws directly implies that the number of generators is also preserved in the dual theory. This is just inverse Noether's theorem implying that those still correspond to continuous symmetries. Hence,
\begin{equation}
G\;\;\mapsto\;\;\widetilde{G}\;:\; \lbrace\; dim(\widetilde{G})=dim(G)\;\;|\;\; \widetilde{G}>\mbox{SO}(1,3) \;\rbrace\;,
\end{equation}
which, thus, may only be
\begin{equation}
\widetilde{G}\;=\;\left\lbrace\begin{array}{l}
\mbox{SO}(1,4)\\
\;\;\;\;\;\mbox{or}\\
\mbox{SO}(2,3)
\end{array}\right.\;.
\end{equation}
$\mbox{SO}(1,4)\cong\mbox{Conf}(3)$ and $\mbox{SO}(2,3)\cong\mbox{Conf}(1,2)$ are the conformal automorphism groups of the conformal compactifications of $\mathbb{R}^{3}$ and $\mathbb{R}^{1,2}$, i.e. of $\mathbb{S}^3$ and $\mathbb{S}^2\times\mathbb{S}^1$, respectively. Actually, those compactifications have antipodal points identified, hence the conformal group for particularly $\mathbb{S}^2\times\mathbb{S}^1$ is $\mbox{SO}(2,3)/\mathbb{Z}_2$. Since the dual space $\mathbb{S}^2\times\mathbb{R}$ is the universal cover of $\mathbb{S}^2\times\mathbb{S}^1$, their conformal groups are locally isomorphic and, therefore,
\begin{equation}
\widetilde{G}\;=\;\mbox{SO}(2,3)\;,
\end{equation}
where SO(1,3) realizes the conformal group on $\mathbb{S}^2$. Hence, since dual coordinates $\widetilde{N}^\mu$ transform under SO(2,3),
\begin{equation}
\mathbb{R}^{1,3}\;\;\mapsto\;\;\left[\mathbb{S}^2\times\mathbb{R}\right]\;,
\end{equation}
which is the conformal equivalence class of the cylinder. At the same time, $\mbox{SO}(2,3)=\mbox{Isom(AdS$_4$)}$, whereas, indeed, the conformal cylinder is the conformal boundary of the universal cover of the compactified AdS$_4$.

\subparagraph{The BMS group}
At this point, we should present the general form of the spin-orbit duality. That is, the map (\ref{MAP}) is really a particular choice among infinite ones,
\begin{equation}
\begin{array}{c}
p\mapsto\lambda\,p\\[5pt]
N\mapsto \frac{1}{\lambda}W\\[5pt]
W\mapsto-\frac{1}{\lambda}N
\end{array}\hspace{0.7cm}\Leftrightarrow\hspace{0.7cm}J\mapsto\star J\;,
\end{equation}
where $\lambda=\lambda(\widetilde{X})$, a function on the dual $\mathbb{S}^2\times\mathbb{R}$ cylinder. Thus, the Lorentz algebra remains intact, which may be verified by applying the generalized map on the analysis in Appendix \ref{App1}. For the rest of the symmetry structure to be preserved, i.e. for $\dot{p}^\mu=0$ to remain true, $\lambda$ must be a function of the $\mathbb{S}^2$ coordinates only, say $\lambda(\theta,\phi)$. Hence, the generalized $\tilde{p}^t=\lambda p^t$ generates a shift,
\begin{equation}
t\;\rightarrow\;t+\lambda(\theta,\phi)\;,
\end{equation}
on the dual space. An analogous situation holds for the remaining symmetries $\tilde{p}^i=\lambda p^i$, after we project them via the Hopf map onto $\mathbb{S}^2$ \cite{footBMS}. Altogether, we obtain an infinite set of, so-called, super-translations on the cylinder. Those, as a semidirect product with the SO(1,3) conformal group on $\mathbb{S}^2$, form the Bondi-Metzner-Sachs (BMS) group \cite{Bondi:1962px}.\\

The BMS group is the group of asymptotic symmetries for asymptotically flat spacetimes, i.e. at null infinity. However, the given set of spaces (that are conformally bounded by null infinity) is larger than that of asymptotically-flat spaces and the BMS group coexists with the conformal group \cite{deBoer:2003vf}. In this setup, this is tantamount to the dual space being the conformal equivalence class of $\mathbb{S}^2\times\mathbb{R}$, the conformal boundary of both $\mathbb{R}^{1,3}$ and AdS$_4$. All facts considered, the appearance of both the BMS and conformal groups justify the identification of the dual cylinder with the $\mathbb{R}^{1,3}$ null infinity.

\section{The Landau levels}
An elementary arena to demonstrate that the map is, in fact, a duality is an electric charge in a uniform magnetic field in $\mathbb{R}^3$ \cite{footHopfIns}. In natural units, its Hamiltonian,
\begin{equation}
\mathcal{H}\;=\;\frac{1}{2m}\left(p^i+A^i(X^j)\right)^2\;,
\end{equation}
where $A^i$ is the vector potential producing the magnetic field $B^i=\epsilon^{ijk}\partial_jA_k$. The spectrum here are the usual Landau levels. The coordinates $X^i$ may as well be regarded as the transverse $N^i$, either by a gauge transformation or by shifting to the rest frame. Then, the dual Hamiltonian,
\begin{equation}
\widetilde{\mathcal{H}}\;=\;\frac{1}{2m}\left(p^i+A^i(\widetilde{X}^j)\right)^2\;,
\end{equation}
where $\widetilde{X}^2=W^2=R^2$, i.e. the dual coordinates $\widetilde{X}\in\mathbb{S}^2$. Since the spin-orbit map is the Hopf map $\mathbb{S}^1\hookrightarrow\mathbb{S}^3\rightarrow\mathbb{S}^2$, the U(1) fiber-connection $A^i$ maps into the potential of a Dirac monopole of minimum charge \cite{foot2} in the center of $\mathbb{S}^2$ \cite{Minami:1979wn}. Of course, $p^i\mapsto p^i$. The spectrum of a charge on a sphere under a magnetic monopole reads \cite{Haldane:1983xm}
\begin{equation}
\widetilde{E}_n\;=\;\frac{1}{2mR^2}\left(n^2+n(2s+1)+s\right)\;,\label{S2spectrum}
\end{equation}
where $s$ is the particle spin. However, this holds on $\mathbb{S}^2$, while the dual space is really the conformal sphere. I.e., the actual dual spectrum must be a kind of scale-invariant limit of (\ref{S2spectrum}), if such a process makes sense.\\

In order to pursue scale invariance we take the limit $R\rightarrow\infty$ holding $B=s/R^2$ (particle density) fixed. In turn, this also implies $s\rightarrow\infty$. This limit makes sense, because the sphere finite effects are suppressed and we should reach a scale-invariant sector of the system, if any. Apparently, such a sector does exist and the process is identified with the \textit{thermodynamic limit},
\begin{equation}
\widetilde{E}_n\rightarrow\omega_c\left(n+\frac{1}{2}\right)\;,
\end{equation}
where $\omega_c=B/m$. The result is indeed scale-invariant and matches precisely the Landau levels of the $\mathbb{R}^3$ bulk.\\

Moreover, the map indicates that the dual vacuum is a fuzzy sphere with $2s+1$ position eigenstates. Again, this too is in perfect agreement with the monopole problem on $\mathbb{S}^2$, where the Lowest Landau Level, $n=0$, is $(2s+1)$-fold degenerate and the $2s+1$ Landau orbitals form a spin-$s$ SO(3) representation, i.e. a fuzzy sphere.

\section{The Ising model}
Expanding the $\mathbb{R}^3$ potential $A^i(X)$, the problem breaks down to a harmonic oscillator coupled to the magnetic field, with a generic form of the Hamiltonian \cite{foot3},
\begin{equation}
\mathcal{H}\;=\;\frac{p^2}{2m}+\frac{1}{2}m\omega_c^2X^2+\omega_LB\cdot L\;,
\end{equation}
for cyclotron and Larmor frequencies $\omega_c$, $\omega_L$, respectively, that are functions of $B$ and $m$. Since the spin-orbit map yields $X^i\mapsto S^i/m$ and $L_i\mapsto S_i$ \cite{foot4}, the dual Hamiltonian,
\begin{equation}
\widetilde{\mathcal{H}}\;=\;\frac{p^2}{2m}+\frac{\omega_c^2}{2m}\,S^2+\omega_LB\cdot S\;.
\end{equation}
This is the Ising model for just one charge. The first term is a kinetic term, whose momenta $p_i$ are conjugate to its $S_i/m$ coordinates \cite{Kopec}; we may discard it by shifting to the rest frame. The second term is a self-interaction \cite{Nareddy}, which, quantum mechanically, is a memory term by virtue of which the new state depends on the old state. The last term is the usual coupling between spin and the transverse magnetic field.\\

Disregarding electron repulsion (e.g. as negligible compared to the magnetic field), we may generalize to $N$ electrons. Then the coordinate is the center-of-mass $X^i=(X_1^i+...+X_N^i)/N$ and the bulk Hamiltonian reads
\begin{equation}
\mathcal{H}\;=\;\sum^N\left(\frac{p^2}{2m}+\frac{1}{2}m\omega_c^2X^2+\omega_LB\cdot L\right)+m\omega_c^2\sum_{a\neq b}^NX_a\cdot X_b\;,
\end{equation}
while its dual counterpart,
\begin{equation}
\widetilde{\mathcal{H}}\;=\;\sum^N\left(\frac{p^2}{2m}+\frac{\omega_c^2}{2m}\,S^2+\omega_LB\cdot S\right)+\frac{\omega_c^2}{m}\sum_{a\neq b}^NS_a\cdot S_b\;.
\end{equation}
The new term is the known inter-site interaction, between all possible spin-lattice sites. That is, not only for next-neighbors (short-range interactions) but for long-range interactions also \cite{Hiley}. This term secretly knows about the scale of the lattice, in the disguise of $\omega_c=B/m$, where $B=s/R^2$ is the particle density. However, the dual system was identified with the thermodynamic limit, where $B$ stays fixed. That is, it is scale-invariant and it is, thus, legitimate to hide the scale inside the (constant) frequency $\omega_c$. Hence, this is an Ising model on fuzzy $\mathbb{S}^2\times\mathbb{R}$ at critical point.\\

Conformal long-range interactions have been studied in \cite{Paulos:2015jfa}. More interestingly, though, conformal invariance on the three-dimensional Ising transition was successfully studied recently in \cite{Zhu:2022gjc}. Their approach was held on $\mathbb{S}^2\times\mathbb{R}$, where $\mathbb{S}^2$ was assumed to be noncommutative, effectively acting as a regulator. The duality presented here is a natural manifestation of that setup, suggesting that bulk electrons in a magnetic field are the hologram of an Ising transition at null infinity.

\section{Holographic fermion vacuum}
All cases studied so far are just simplified, quantum-mechanical versions of quantum electrodynamics. Nonetheless, because the Hopf map acts non-trivially on the gauge field, we first attempt looking into the vacuum of free fermions,
\begin{equation}
S\;=\;\int_{\mathbb{R}^{1,3}}\;\bar{\psi}\left(i\slashed{\partial}-m\right)\psi\;.\label{FreeSpinors}
\end{equation}
The expectations on the transformation rule of this action are the following. First, since $p^\mu\mapsto p^\mu$, the kinetic term stays invariant. Secondly, because of the dual global symmetry $\widetilde{G}=\mbox{SO}(2,3)$, there should be no mass term in the dual action. On the contrary, since the dual space is $\mathbb{S}^2\times\mathbb{R}$, a spin connection should, somehow, emerge \cite{foot5}.

\subparagraph{Generalized field coordinates}
Since coordinates transform, $N^\mu\mapsto W^\mu$, the expectation that the mass term transforms too is partly because $\bar{\psi}\psi$ is the fermion probability density, the analog of a coordinate operator in field theory. In fact, this analogy can be made formal. That is, we may use the generalized momenta $\Pi_\mu=i\partial_\mu\psi$ and $\bar{\Pi}_\mu=i\partial_\mu\bar{\psi}$ to define the generalized field coordinates $\partial\mathcal{L}/\partial\bar{\Pi}_\mu=\gamma^\mu\psi$ and $\partial\mathcal{L}/\partial\Pi_\mu=\bar{\psi}\gamma^\mu$, respectively. Normalizing as
\begin{equation}
\mathsf{\Psi}^\mu:=\frac{\gamma^\mu\psi}{2\sqrt{-p^2}}\hspace{0.5cm}\mbox{and}\hspace{0.5cm}\overline{\mathsf{\Psi}}^\mu:=-\frac{\bar{\psi}\gamma^\mu}{2\sqrt{-p^2}}\;,
\end{equation}
to adjust length dimensions, these coordinates make sense because, first, $\overline{\Psi}^\mu\Psi_\mu=\bar{\psi}\psi/m^2$ is still the probability density. Secondly,
\begin{equation}
[\overline{\mathsf{\Psi}}^\mu,\mathsf{\Psi}^\nu]\;=\;-\frac{i\bar{\psi}\,\mathbf{S}^{\mu\nu}\psi}{p^2}\;,
\end{equation}
analogous to coordinates' noncommutative algebra (\ref{BulkFuzzy}). In Appendix \ref{App4} we show that we may even extract a spacelike coordinate $\mathsf{N}^\mu$, the analog of $N^\mu=L^{\mu\nu}p_\nu/p^2$,
\begin{equation}
\mathsf{N}^\mu:=\frac{\mathbf{S}^{\mu\nu}\partial_\nu\psi}{p^2}\hspace{0.5cm}\mbox{and}\hspace{0.5cm}\overline{\mathsf{N}}^\mu:=\frac{\partial_\nu\bar{\psi}\,\mathbf{S}^{\nu\mu}}{p^2}\;.
\end{equation}
Hence, we may define an analog of orbital angular momentum,
\begin{equation}
\mathfrak{L}^{\mu\nu}\;:=\;\,\frac{\mathbf{S}^{\mu\rho}\partial_\rho}{p^2}\partial_\nu\;-\;\frac{\mathbf{S}^{\nu\rho}\partial_\rho}{p^2}\partial_\mu\;,
\end{equation}
acting on Dirac spinors, defining, in turn, the total angular momentum generator,
\begin{equation}
\mathfrak{J}^{\mu\nu}\;=\;\mathfrak{L}^{\mu\nu}\;+\;\mathfrak{S}^{\mu\nu}\;,
\end{equation}
where $\mathfrak{S}^{\mu\nu}=\mathbf{S}^{\mu\nu}/2$ is the spin part \cite{foot6}. This does satisfy the Lorentz algebra and, therefore, is one of its legitimate representations. In fact, this is the formal contact point we were craving: the spin-orbit map is the statement that $J^{\mu\nu}\mapsto\star J^{\mu\nu}$ or, in this representation, $\mathfrak{J}^{\mu\nu}\mapsto\star \mathfrak{J}^{\mu\nu}$. This is tantamount to the map
\begin{equation}
\begin{split}
\mathsf{N}^\mu\;\;\;&\mapsto\;\;\;\mathsf{W}^\mu\\
\mathsf{W}^\mu\;\;\;&\mapsto\;\;-\mathsf{N}^\mu
\end{split}
\end{equation}
where $\mathsf{W}^\mu=i\,\overrightarrow{\partial}_\nu\star\mathbf{S}^{\mu\nu}\psi/p^2$ is the Pauli-Lubanski (coordinate) vector in position space. This is the field-theoretic equivalent of the spin-orbit duality map.\\

Hence, it may be that the kinetic term stays invariant, but the mass term transforms,
\begin{equation}
m\,\bar{\psi}\psi\;\;\;\mapsto\;\;\;\pm\left(\frac{i}{4}\,\bar{\lambda}_\pm\,\gamma^\alpha \,{\omega_\alpha}^{ab}\sigma_{ab}\,\lambda_\pm\right)\;,
\end{equation}
where $\lambda_\pm$ are (two-component) Dirac spinors in three dimensions and ${\omega_\alpha}^{ab}$ is the spin connection on $\mathbb{S}^2\times\mathbb{R}$, while $\alpha$ and $a,b$ refer to three-dimensional curved and flat indices, respectively. Under the dimensional reduction, $N_f$ four-dimensional Dirac spinors $\psi$ reduce to $2N_f$ three-dimensional $\lambda_\pm$. The $\pm1$ factor is the charge under which $\lambda_\pm$ couple to the $\mathbb{S}^2$ curvature, a relic of the four-dimensional chirality. The derivation is in Appendix \ref{App5}.\\

Let us appreciate this result: the mass term transforms, as it must for a dual conformal field theory (CFT), into the spin connection exactly needed for spinor transport on the (conformal) cylinder. The dual action,
\begin{equation}
\widetilde{S}\;=\;\int_{\tiny\mbox{$[\mathbb{S}^2\times\mathbb{R}]$}}\bar{\lambda}_\pm\left(i\slashed{\partial}\pm\frac{i}{4}\gamma^\alpha{\omega_\alpha}^{ab}\,\sigma_{ab}\right)\lambda_\pm\;.\label{DualQED}
\end{equation}
implying that $N_f$ massive fermions in $\mathbb{R}^{1,3}$ are the hologram of $2N_f$ massless ones at  $[\mathbb{S}^2\times\mathbb{R}]$ null infinity. Note, also, that an even number of flavors guarantees parity and time-reversal symmetry conservation \cite{Borokhov:2002ib}.

\section{The path integral}
As we must, in order to test the proposed duality, we look into the vacuum of free fermions, for both the massive $\mathbb{R}^4$ theory and the massless one at null infinity. In $\mathbb{R}^4$, a free massive Dirac fermion has a path integral,
\begin{equation}
Z_4\;=\;\exp{-\frac{V_4\,m^4}{(4\pi)^2}\left(\log{\frac{\mu^2}{m^2}}\;+\;\mbox{finite}\right)}\;,
\end{equation}
where $V_4$ is the spacetime volume and $\mu$ the energy renormalization scale, a simple exercise kept in Appendix \ref{App6}. On the $\mathbb{S}^2\times\mathbb{R}$ cylinder the situation is different: Dirac modes on $\mathbb{S}^2$ act as an effective mass along $\mathbb{R}$. Both the $\mathbb{S}^2$ modes and the $\mathbb{R}$ momentum have to be regulated, a process held in Appendix \ref{App7}. At the end, two massless fermions exhibit a vacuum,
\begin{equation}
Z_3\;=\;\exp\left\lbrace-\frac{V_3\,R^{-3}}{(4\pi)^{\frac{3}{2}}}\left(\log\frac{R^2}{\rho^2}+\mbox{finite}\right)\right\rbrace\;,
\end{equation}
where $V_3$ is the cylinder volume and $\rho$ a length renormalization scale. However, the cylinder radius is $R=1/m$ \cite{foot9}, implying
\begin{equation}
Z_3\;=\;\exp\left\lbrace-\frac{V_3\,m^3}{(4\pi)^{\frac{3}{2}}}\left(\log\frac{\mu^2}{m^2}+\mbox{finite}\right)\right\rbrace\;,
\end{equation}
where we changed length to an energy renormalization scale. As said, dual space is actually the conformal class $[\mathbb{S}^2\times\mathbb{R}]$, which should indicate the presence of a conformal anomaly. In three dimensions, though, there can be no bulk anomaly, albeit a boundary contribution is possible. Since the cylinder has no boundary, no anomaly exists at this point of the conformal class \cite{foot10}.\\

On the other hand, [$\mathbb{S}^2\times\mathbb{R}$] is the conformal boundary of (the conformally compactified) AdS$_4$. Actually, it is the boundary of the universal cover of AdS$_4$, where time is decompactified on the real line and closed timelike curves are naturally avoided. A massive fermion in $\mathbb{R}^4$ that is (spin-orbit) dual to two massless fermions at null infinity may only be (AdS/CFT) dual to a massive fermion (of the same $\mathbb{R}^{1,3}$ mass) in this enclosed AdS$_4$. This may be regarded as a massive mode of the AdS superstring in the limit where coupling and worldsheet effects are suppressed. Regardless, it is the only option producing a vacuum equal to the ones already given \cite{footVasiliev}.\\

\begin{figure}[t!]
    {{\includegraphics[width=\linewidth]{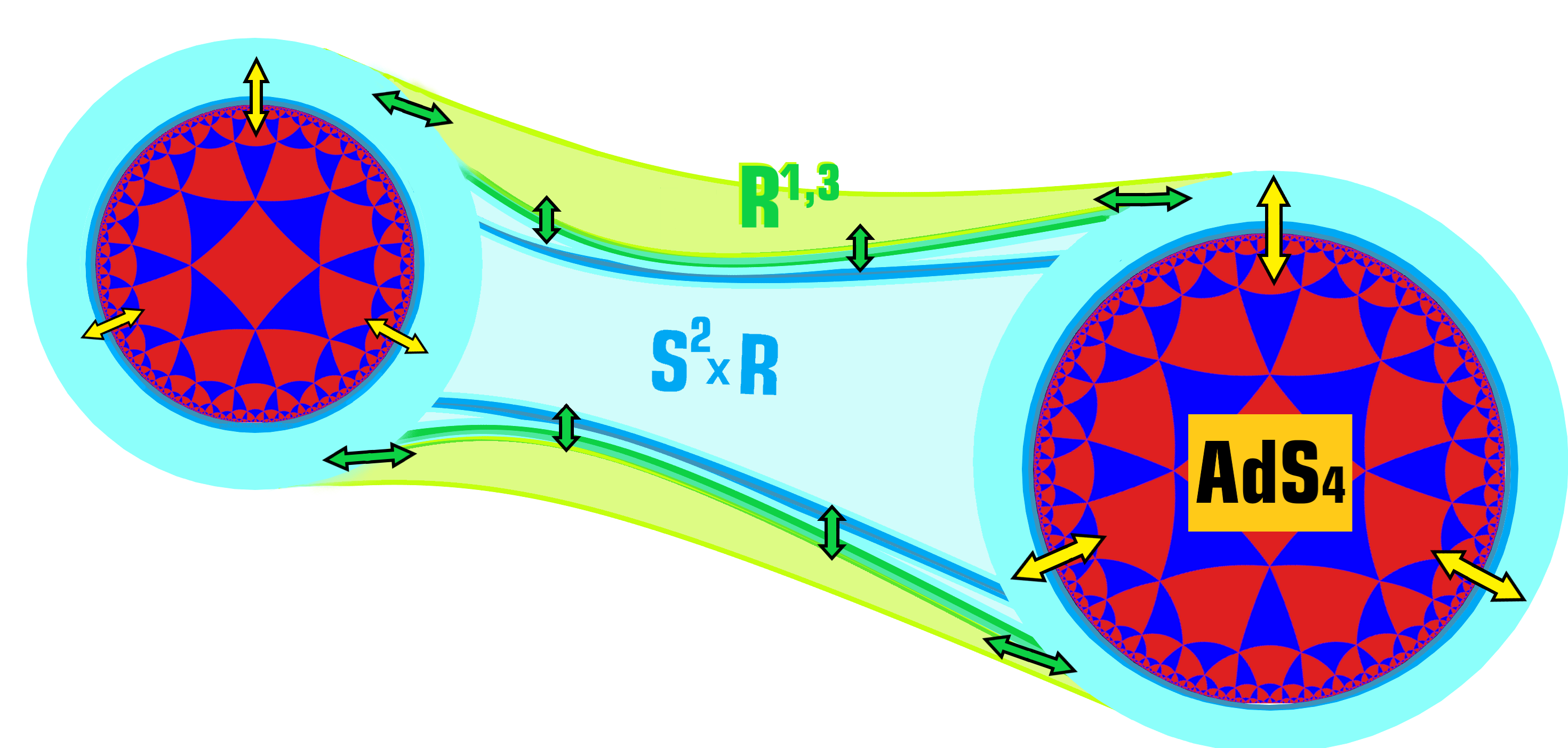} }}%
\caption{This cartoon is a superposition of both dualities. Minkowski space (green) maps to the cylinder (blue) through the spin-orbit duality (green arrows). The cylinder encloses the universal cover of compactified AdS$_4$ (red-blue), those two connected through the AdS/CFT duality (yellow arrows). $\mathbb{S}^2\times\mathbb{R}$ is a lightlike boundary of $\mathbb{R}^{1,3}$, a.k.a. null infinity, and, at the same time, a timelike boundary of the enclosed AdS$_4$. In that sense, one holography nests inside the other.}
 \label{SUPER}
\end{figure}

Indeed, while AdS is boundary-less and the corresponding Dirac spectrum is continuous \cite{Camporesi:1995fb}, (the universal cover of) its compactification exhibits boundary conditions on its conformal boundary which imply discrete (and regular) Dirac modes \cite{Cotaescu:1998ts}. Hence, for a Dirac fermion of mass $m$,
\begin{equation}
Z_A\;=\;\exp\bigg\lbrace-\frac{1}{4}\frac{V_{A}\,(m^2+R^{-2})^{2}}{(4\pi)^{2}}\log\left(\frac{\mu^2}{m^2+R^{-2}}\right)\bigg\rbrace\;,
\end{equation}
plus finite terms, where $V_A$ is the AdS$_4$ volume and $R$ is the AdS radius inherited from the boundary $\mathbb{S}^2\times\mathbb{R}$. The derivation sits in Appendix \ref{App8}. As expected, the logarithmic divergence depends on both scales present. However, $R=1/m$ and, therefore, there is really only one scale,
\begin{equation}
Z_A\;=\;\exp\left\lbrace-\frac{V_A\,m^4}{(4\pi)^{2}}\left(\log\frac{\mu^2}{m^2}+\mbox{finite}\right)\right\rbrace\;.
\end{equation}
Hence, all three path integrals match exactly,
\begin{equation}
Z_4\;=\;Z_3\;=\;Z_A
\end{equation}
up to the dimensionless $d$-volume form $V_dm^d/(4\pi)^{d/2}$. Under both dualities, which are superposed in Figure \ref{SUPER}, those formally infinite volumes are conformal immersions or submersions of each other.

\section{Epilogue}
In summary, the proposed picture is the following. Spin-orbit duality maps a massive $\mathbb{R}^{1,3}$ theory into a massless one on the (null) conformal boundary $\mathbb{S}^2\times\mathbb{R}$. The dual space is actually the conformal class $[\mathbb{S}^2\times\mathbb{R}]$, the (timelike) conformal boundary of AdS$_4$. Hence, massive fields in Minkowski space are the hologram of massless ones at null infinity, which, in turn, are the hologram of superstring modes inside an interior AdS$_4$ universe. This suggests a double holographic bridge,
\begin{equation*}
\left[\begin{array}{c}
\mathbb{R}^{1,3}\\
\mbox{QFT} 
\end{array}\right]\xrightleftharpoons[\mbox{\scriptsize duality}]{\mbox{\scriptsize spin-orbit}}
\left[\begin{array}{c}
[\mathbb{S}^2\times\mathbb{R}]\\
\mbox{CFT}
\end{array}\right]\xrightleftharpoons[\mbox{\scriptsize duality}]{\mbox{\scriptsize AdS/CFT}}
\left[\begin{array}{c}
\mbox{AdS$_4$}\times\mathcal{M}^6\\
\mbox{superstring}
\end{array}\right]
\end{equation*}\\
which, at this stage, manifests through the matching of all three free path integrals, in $\mathbb{R}^4$ and $\mathbb{S}^2\times\mathbb{R}$ and AdS$_4$. This state of affairs is what we call nested holography.\\

%The duality, moreover, is a realization of the Hopf map. Hence, a U(1) fiber-connection in $\mathbb{R}^{1,3}$ maps to the gauge field of a Dirac monopole on the $\mathbb{S}^2\times\mathbb{R}$ cylinder.\\

Mapping free massive $\mathbb{R}^{1,3}$ spinors, i.e. (\ref{FreeSpinors}), into massless ones at null infinity, i.e. (\ref{DualQED}), and matching their path integrals represents a first step towards an exact Minkowski/CFT correspondence. Work in this direction is known as the flat-space holography program \cite{deBoer:2003vf,Arcioni:2003xx}, Celestial \cite{Cheung:2016iub} and Carrollian \cite{Dappiaggi:2005ci} holography. However, we underline three important differences. First, spin-orbit holography maps $\mathbb{R}^{1,3}$ fields directly onto null infinity, it does not identify fields' asymptotic behavior with a boundary CFT. Secondly, and in that respect, it maps massive configurations, not just massless ones. Thirdly, it maps ordinary quantum field theory, not quantum gravity, to a CFT.\\

The free fermion vacuum sets the foundations but, otherwise, we are interested in the holography of a fully interacting theory. Spin-orbit holography realizes the Hopf map which, as said, acts non-trivially on the U(1) gauge theory, nevertheless confessing some of its key properties on $[\mathbb{S}^2\times\mathbb{R}]$. One is the dimension of the gauge coupling, $[g]=m^{1/2}$, which yields super-renormalizability. Another is the fate of the $\mathbb{R}^{1,3}$ gauge field, which reduces on the cylinder into a gauge field and a scalar field $\phi$ which couples to fermions via a Yukawa coupling $g\bar{\lambda}\phi\lambda$. In three dimensions, a Chern-Simons term may also exist and, as we will present in future work, this is indeed the case. Ultimately, the final theory (or a limit thereof) should be conformal, since a CFT is what we anticipate at null infinity. Obtaining this CFT will then let us look for a proper (AdS/CFT) superstring dual in the interior AdS$_4$ space \cite{Maldacena:1997re}.\\

In regard to the interior AdS$_4$ vacuum, we also note that there is a natural compliance with the AdS distance conjecture \cite{Lust:2019zwm}. That is, AdS$_4$ inherits its radius from the $\mathbb{R}^{1,3}$ fermion mass, $R=1/m$, fixing the vacuum across the moduli space of varying radii. This implies that the infinite Kaluza-Klein tower of states, which is controlled by the cosmological constant $\Lambda$ (which, in turn, is set by the AdS radius), does not get light and, therefore, the effective field theory does not breakdown. In fact, the compactification mass $m_{\tiny\mbox{KK}}\sim\abs{\Lambda}^{1/2}=m$. In that sense, the interior AdS vacuum is guaranteed to be part of the superstring landscape, as it should for any sensible dual $\mathbb{R}^{1,3}$ vacuum.\\

\begin{figure}[t!]
    {{\includegraphics[width=\linewidth]{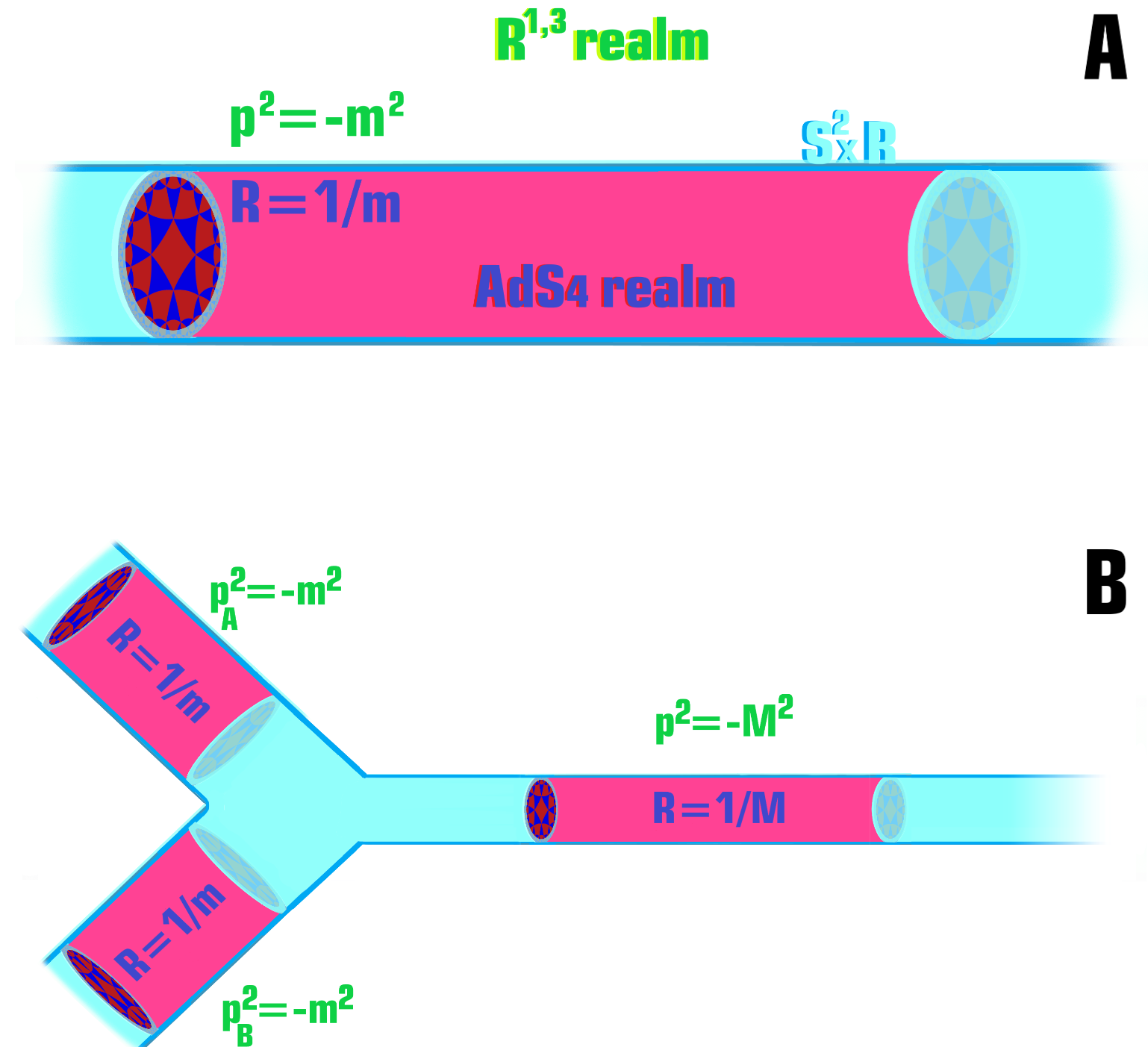} }}%
\caption{The $\mathbb{R}^{1,3}$ and AdS$_4$ realms as a unified framework.}
 \label{REALM}
\end{figure}

We conclude by presenting a unified framework. Consider the worldline of a massive fermion in $\mathbb{R}^{1,3}$. Relativistically, as explained, this actually is a world-tube with $\mathbb{S}^2\times\mathbb{R}$ topology enclosing all possible pseudo-worldlines. We call this basic view the $\mathbb{R}^{1,3}$ realm. Nested holography, then, through $\mathbb{S}^2\times\mathbb{R}$ null infinity, takes us to an interior AdS$_4$ universe, whose radius $R=1/m$, the particle's Compton wavelength. This view is the AdS$_4$ realm. Identifying the Minkowski world-tube with null infinity in virtue of their common  $\mathbb{S}^2\times\mathbb{R}$ topology, both realms are superposed in Figure \ref{REALM}A. Because we calculated path integrals in the free limit, effectively $g_s\rightarrow0$ and $R/l_s\gg1$. That is, coupling and worldsheet effects are suppressed. Hence, although $R$ is small, $R/l_P\gg1$ (since $l_P^4=g_sl_s^4$) implying that quantum gravity too is suppressed in the AdS$_4$ realm. All those facts form an analogy to an observer in the $\mathbb{R}^{1,3}$ realm, who just sees free (or asymptotic) states of elementary particles.\\

In an interacting theory, however, $g_s$ and $l_s$ can be anything. In the $\mathbb{R}^{1,3}$ realm, multiple particle worldlines merge into new ones of larger four-momentum. In the AdS$_4$ realm, the radius becomes small (smaller than that of free or asymptotic states), presumably $R/l_s\lesssim1$ and $R/l_P\lesssim1$, where worldsheet and gravitational effects become important. A small radius for the $\mbox{AdS}_5\times\mathbb{S}^5$ superstring has been addressed in \cite{Gopakumar:1998ki}. This is all illustrated in Figure \ref{REALM}B. Equally, a small AdS radius could, alternatively, be obtained by considering the large mass of a black hole in the $\mathbb{R}^{1,3}$ realm, where quantum gravity is indeed anticipated.\\

Overall, the present work offers two key insights. One is the spin-orbit duality, which introduces a Minkowski/CFT correspondence. The other is that the AdS/CFT duality is placed into a larger context, the CFT being a common ground between both dualities. Said differently, every massive theory in $\mathbb{R}^{1,3}$ is conjectured to be dual to an AdS/CFT setup living at its null infinity. This implies that, in fact, it is the AdS superstring that associates with fields in flat space.

\subparagraph{Acknowledgments}
I thank C. Nunez and D. Giataganas for their comments, Kate and Hector Liatsos for our conversation. I am grateful to Fani Christidi for her insights on the concepts introduced. This is dedicated to the memory of Eleni Topaloudi.

%\newpage

\appendix

\section{A Lorentz algebra automorphism}\label{App1}
The Lorentz algebra is preserved, a fact that is anticipated from the preservation of the conservation law $\dot{J}=0$: inverse Noether's theorem implies that this still corresponds to a continuous symmetry. Geometrically, $J\mapsto\star J$ is a swap between the hyperplanes of rotations and boosts, i.e. the topological invariance
\begin{equation}
\mathbb{RP}^3\times\mathbb{R}^3\;\;\mapsto\;\;\mathbb{R}^3\times\mathbb{RP}^3\;.
\end{equation}
All this should be expected, since the Hodge star is a linear map that just shifts orthonormal basis. A detailed proof of the invariance of the Lorentz algebra under spin-orbit duality is held in \cite{Filippas:2022sna}. An alternate proof is the following. Considering the Poincar\`e generators $\lbrace\mathbf{J,P}\rbrace$, their possible compositions are
\begin{equation}
\mathbf{W}^\mu:=\frac{\star(\mathbf{J}^{\mu\nu}\wedge\mathbf{P}_\nu)}{\mathbf{P}^2}\hspace{0.7cm}\mbox{and}\hspace{0.7cm}\mathbf{N}^\mu:=\frac{\mathbf{J}^{\mu\nu}\cdot\mathbf{P}_\nu}{\mathbf{P}^2}
\end{equation}
where we normalized over to length dimensions. $\mathbf{W}$ is the Pauli-Lubanski generator and $\mathbf{N}$ is the algebraic analog of the spacelike position $N^\mu$, the latter defining the orbital-angular-momentum generator $\mathbf{L}^{\mu\nu}=\mathbf{N}^\mu\wedge\mathbf{P}^\nu$. Together, they reconstruct the Lorentz algebra,
\begin{equation}
[\mathbf{W}^\mu,\mathbf{W}^\nu]=\frac{\mathbf{J}^{\mu\nu}}{\mathbf{P}^2}\;,\;[\mathbf{W}^\mu,\mathbf{N}^\nu]=\frac{\star\mathbf{J}^{\mu\nu}}{\mathbf{P}^2}\;,\;[\mathbf{N}^\mu,\mathbf{N}^\nu]=-\frac{\mathbf{J}^{\mu\nu}}{\mathbf{P}^2}\;,
\end{equation}
where $\mathbf{W}$ generates SO(3) and $\mathbf{N}$ boosts. Then, this is left invariant by the map
\begin{equation}
\begin{array}{c}
\mathbf{N}\mapsto\mathbf{W}\\[6pt]
\mathbf{W}\mapsto-\mathbf{N}
\end{array}\hspace{0.7cm}\Leftrightarrow\hspace{0.7cm}\mathbf{J}\mapsto\star \mathbf{J}
\end{equation}
up to an orthonormal change of basis on generators $\mathbf{J}$.

\section{Noncommutative Minkowski spacetime}\label{App2}
The Lorentz algebra is an exterior algebra on coordinates and momentum and, as such, involves only the spacelike $N^\mu$. Hence, a generator $\mathbf{A}$, representative of the timelike coordinate $A^\mu$, cannot be built out of Poincar\`e generators. Therefore, we introduce the generator $\mathbf{X}$, representative of the coordinate $X^\mu$, with the sole requirement that it is conjugate to $\mathbf{P}$, that is $[\mathbf{X},\mathbf{P}]=\eta\mathbf{1}$. Then $\mathbf{A}$ is defined by $\mathbf{A}^\mu=\mathbf{X}^\mu-\mathbf{N}^\mu=(\mathbf{X}\cdot\mathbf{P})\mathbf{P}^\mu/\mathbf{P}^2$, implying $[\mathbf{A}^\mu,\mathbf{P}^\nu]=\mathbf{P}^\mu\mathbf{P}^\nu/\mathbf{P}^2$, as it should. As with $\mathbf{N}$, the order of the generators is irrelevant. All this gives
\begin{equation}
[\mathbf{X}^\mu,\mathbf{X}^\nu]\;=\;[\mathbf{X}^\mu,\mathbf{X}^\alpha]\;\frac{\mathbf{P}_\alpha\mathbf{P}^\nu}{\mathbf{P}^2}\;-\;[\mathbf{X}^\nu,\mathbf{X}^\alpha]\;\frac{\mathbf{P}_\alpha\mathbf{P}^\mu}{\mathbf{P}^2}\;-\;\frac{\mathbf{S}^{\mu\nu}}{\mathbf{P}^2}\;\;.
\end{equation}
A naive solution is $[\mathbf{X},\mathbf{X}]=-\mathbf{J}/\mathbf{P}^2$, but this is discarded because the Jacobi identity implies that $[\mathbf{X},\mathbf{X}]$ commutes with $\mathbf{P}$, contradicting the proposal. Thus, the unique solution is
\begin{equation}
[\mathbf{X}^\mu,\mathbf{X}^\nu]\;=\;-\frac{\mathbf{S}^{\mu\nu}}{\mathbf{P}^2}\;.
\end{equation}

\section{From $\mathbb{R}^{1,3}$ to $\mathbb{S}^2\times\mathbb{R}$}\label{App3}
Since the map becomes trivial on a certain radius, it effectively acts on the domain $\mathbb{R}^3\setminus\lbrace0\rbrace$, endowed with the metric $\dd s^2=\dd n^i\dd n_i/n^in_i$. Algebraically, the subtracted center renders this space a regular submanifold of Minkowski space \cite{footQin}. In turn, $\mathbb{R}^3\setminus\lbrace0\rbrace$ is isomorphic to $\mathbb{S}^2\times\mathbb{R}$. In fact, $\phi: \mathbb{S}^2\times\mathbb{R}\rightarrow\mathbb{R}^3\setminus\lbrace0\rbrace$ is a smooth map $\phi(\chi,r)=e^r\chi$, with an inverse $\phi^{-1}(n)=(n/\abs{n},\log\abs{n})$ and $n\in\mathbb{R}^3\setminus\lbrace0\rbrace$. Now, given the two projections that can be defined on $\mathbb{S}^2\times\mathbb{R}$, namely $\Pi_1: \mathbb{S}^2\times\mathbb{R}\rightarrow\mathbb{S}^2$ and $\Pi_2: \mathbb{S}^2\times\mathbb{R}\rightarrow\mathbb{R}$, the map $\phi$ defines an immersion $\pi_a=\Pi_a\circ\phi^{-1}$ for $a=1,2$. The duality projects on just a point along $\mathbb{R}$, i.e. $\abs{n}=\abs{W}=\rho$, which redefines $\phi^{-1}_\rho=(n/\abs{n},\rho)$ and produces the overall map $\pi_3:=\pi_1\circ\pi_2$,
\begin{equation}
n\;\;\mapsto\;\;\tilde{n}\;=\;\pi_3(n)\;=\;\frac{n}{\abs{n}}\rho
\end{equation}
Hence, this a conformal immersion $\mathbb{R}^3\setminus\lbrace0\rbrace\rightarrow\mathbb{S}^2$.

\section{Spacelike field coordinate}\label{App4}
We may extract a spacelike coordinate $\mathsf{N}^\mu$, the analog of $N^\mu$, by considering the (position-space) projector $A_{\mu\nu}=i^2\overleftarrow{\partial}_\mu\overrightarrow{\partial}_\nu/p^2$,
\begin{equation}
\overline{\mathsf{\Psi}}\cdot \mathsf{N}\;=\;\overline{\mathsf{\Psi}}\cdot\mathsf{\Psi}-\overline{\mathsf{\Psi}}\cdot (A\cdot\mathsf{\Psi})\;=\;\frac{\bar{\psi}\psi}{m^2}-\frac{1}{4}\frac{\bar{\psi}\psi}{m^2}\;=\;\frac{3}{4}\frac{\bar{\psi}\psi}{m^2}\;,
\end{equation}
where we used the equation of motion. Notice that the numerical factors naturally decompose into timelike and spacelike degrees of freedom. We can obtain the exact same result just by decomposing $i\gamma^\mu\gamma^\nu=-2\mathbf{S}^{\mu\nu}-i\eta^{\mu\nu}$ on the Dirac equation, deducing an explicit expression,
\begin{equation}
\mathsf{N}^\mu:=\frac{\mathbf{S}^{\mu\nu}\partial_\nu\psi}{p^2}\hspace{0.5cm}\mbox{and}\hspace{0.5cm}\overline{\mathsf{N}}^\mu:=\frac{\partial_\nu\bar{\psi}\,\mathbf{S}^{\nu\mu}}{p^2}\;.
\end{equation}
Given $N^\mu=J^{\mu\nu}p_\nu/p^2=L^{\mu\nu}p_\nu/p^2$, this form makes sense because it is $\mathfrak{su}(2)\times\mathfrak{su}(2)$ that rotates spinors.

\section{Mass to spin connection}\label{App5}
Extracting the timelike coordinate out of the mass term and merging it into the kinetic term, the action,
\begin{equation}
S=\int_{\mathbb{R}^{1,3}}i\bar{\psi}\slashed{\partial}\psi-m\bar{\psi}\psi\;=\;\int_{\mathbb{R}^{1,3}}\frac{3}{4}i\bar{\psi}\slashed{\partial}\psi-\frac{3}{4}m\bar{\psi}\psi
\end{equation}
where now the mass term represents only the spacelike projection $\bar{\mathsf{\Psi}}\cdot\mathsf{N}$. The overall numerical factor is irrelevant and may be set to unity by a field redefinition; this redefinition is really the choice of isolating only the spatial degrees of freedom into the density $\bar{\psi}\psi$. Therefore,
\begin{equation}
m\,\bar{\psi}\psi\;\;\rightarrow\;\; \;m^3\,\overline{\mathsf{\Psi}}\cdot\mathsf{N}\;.
\end{equation}
Hence, the mass term transforms as
\begin{equation}
\overline{\mathsf{\Psi}}\cdot\mathsf{N}=\frac{1}{2}\left(\overline{\mathsf{\Psi}}\cdot\mathsf{N}+\overline{\mathsf{N}}\cdot\mathsf{\Psi}\right)\;\;\mapsto\;\;\frac{1}{2}\left(\overline{\mathsf{\Psi}}\cdot\mathsf{W}+\overline{\mathsf{W}}\cdot\mathsf{\Psi}\right)\;.
\end{equation}
After integrating by parts, in the original field language,
\begin{equation}
\begin{split}
m\,\bar{\psi}\psi\;\;\mapsto\;\;&\frac{\bar{\psi}\gamma^\mu}{2}(\frac{i}{2}\epsilon_{\mu\nu\rho\sigma}(\partial^\nu\mathbf{S}^{\rho\sigma})\psi)\\[10pt]
=&i\,\frac{\bar{\psi}\gamma^\mu}{2}(\partial^\nu\frac{\sigma_{\mu\nu}}{2})\gamma_5\psi\;,
\end{split}
\end{equation}
where we used the duality identity $\frac{1}{2}\epsilon_{\mu\nu\rho\sigma}\sigma^{\rho\sigma}=\gamma_5\sigma_{\mu\nu}$, while the spinor representation of Lorentz generators $\mathbf{S}_{\mu\nu}=\frac{\sigma_{\mu\nu}}{2}=-\frac{i}{4}[\gamma_\mu,\gamma_\nu]$. Using the projectors $P_{\pm}=(1\pm\gamma^5)/2$ (and the fact that there is an odd number of Dirac matrices), we express the dual quantity in terms of (the two) three-dimensional spinors $\lambda_\pm$ (see \cite{Angelone:2022tet} for a related discussion). The latter breaks into two terms $I_\pm$, one for each spinor $\lambda_\pm$,
\begin{equation}
\begin{split}
I_\pm\;=\;&\pm\left(\frac{i}{4}\bar{\lambda}_\pm\gamma^\alpha(\partial^\beta\sigma_{\alpha\beta})\lambda_\pm\right)\\
=&\pm\left(\frac{i}{4}\bar{\lambda}_\pm\gamma^\alpha\left[ e^\mu_\alpha e^\nu_\beta\partial^\beta(e^\gamma_\mu e^\delta_\nu e_\gamma^a e_\delta^b)\right]\sigma_{ab}\lambda_\pm\right)\;,
\end{split}
\end{equation}
where $\alpha,\beta$ and $a,b$ refer to three-dimensional curved and flat indices, respectively. The components $e^\mu_\alpha$ are the (holographic) map between the three-dimensional and the four-dimensional space, while $e^\alpha_a$ the usual vielbein in three dimensions. After a long but straightforward calculation,
\begin{equation}
I_\pm=\pm\left(\frac{i}{4}\bar{\lambda}_\pm\gamma^\alpha \left[-e_\beta^b\,\Gamma^\beta_{\gamma\alpha}\,e^{\gamma a}+e_\alpha^a\nabla^\gamma e^b_\gamma+e_\beta^b\,\partial^\beta e^a_\alpha\right]\sigma_{ab}\lambda_\pm\right)\;,
\end{equation}
where the second term vanishes, since $\nabla^\sigma e^b_\sigma=0$ for a torsionless connection. Hence,
\begin{equation}
I_\pm\;=\;\pm\left(\frac{i}{4}\bar{\lambda}_\pm\gamma^\mu \left[e_\nu^b\,\nabla^\nu e^a_\mu\right]\sigma_{ab}\lambda_\pm\right)\;,
\end{equation}
which is the coupling to the spin connection on $\mathbb{S}^2\times\mathbb{R}$,
\begin{equation}
I_\pm\;=\;\pm\left(\frac{i}{4}\bar{\lambda}_\pm\gamma^\mu \,{\omega_\mu}^{ab}\,\sigma_{ab}\lambda_\pm\right)\;.
\end{equation}

\section{Path integral in $\mathbb{R}^4$}\label{App6}
The Euclidean effective action for a Dirac fermion is
\begin{equation}
S_{\text{eff}} = -\log Z_4 = -\frac{1}{2}\int_{\mathbb{R}^4} \text{Tr} \log(p^2 - m^2)\;,
\end{equation}
where the trace is both over spinor and momentum space. Continuing spacetime dimensions to \( d = 4 - \epsilon \),
\begin{equation}
S_{\text{eff}} = -2V_4 \mu^{4-d} \int \frac{d^d p}{(2\pi)^d} \log(p^2 - m^2).
\end{equation}
where $V_4$ is the space volume. To get rid of the logarithm, we may differentiate (and, at the end, integrate) with respect to $m^2$. Performing the integral,
\begin{equation}
S_{\text{eff}} = -2V_4 \mu^{4-d} \left\lbrace\frac{(m^2)^{d/2}}{(4\pi)^{d/2}d/2}\Gamma\left( 1 - \frac{d}{2} \right)+\mbox{const.}\right\rbrace.
\end{equation}
Expanding $d = 4 - \epsilon$ and renormalizing, the finite part of the effective action is
\begin{equation}
S_{\text{eff}} = \frac{V_4\,m^4}{16 \pi^2} \left\lbrace \log\left( \frac{\mu^2}{m^2} \right) + \mbox{finite} \right\rbrace\;.
\end{equation}

\section{Path integral on $\mathbb{S}^2\times\mathbb{R}$}\label{App7}
The Euclidean Dirac operator $\slashed{\mathcal{D}} = \slashed{\mathcal{D}}_{\mathbb{S}^2} + \gamma^t \partial_t$ implies
\begin{equation}
Z_3=\det\slashed{\mathcal{D}}=\prod_k\lambda_k=e^{-\frac{1}{2}\zeta'_{\mbox{\tiny$\mathcal{D}^2$}}(0)-\frac{1}{2}\zeta(0)_{\mbox{\tiny$\mathcal{D}^2$}}\log\mu^2}
\end{equation}
where
\begin{equation}
\zeta_{\mbox{\tiny$\mathcal{D}^2$}}(s)=\sum_{l,p}\lambda_{l,p}^{-2s}=\sum_{l,p}N_l\left[\left(\frac{l+1}{R}\right)^2+p_t^2\right]^{-s}
\end{equation}
is the spectral zeta function on $\mathcal{D}^2$. The first term is the angular momentum eigenvalues (of $N_l=2(l+1)$ degeneracy), corresponding to spherical harmonics on $\mathbb{S}^2$ \cite{Bar}, while the second is the momentum along $\mathbb{R}$. Thus, a massless fermion on $\mathbb{S}^2\times\mathbb{R}$ may be seen as a one-dimensional massive particle along $\mathbb{R}$, with mass $\abs{\frac{l+1}{R}}$. While the zeta function regulates the infinite eigenvalues, a subtraction scheme is also necessary. This would be the heat kernel, naturally connected to the zeta function,
\begin{equation}
\zeta_{\mbox{\tiny$\mathcal{D}^2$}}(s)=\frac{1}{\Gamma(s)}\int_0^{\infty}\dd\sigma\sigma^{s-1}K(\sigma)
\end{equation}
where
\begin{equation}
K(\sigma)=\int\dd^3x\,\Tr e^{-\sigma\mathcal{D}^2}=2\sum_{l,p}\int\dd^dx\,\frac{e^{-\sigma\lambda_{l,p}^2}}{(4\pi\sigma)^{\frac{d}{2}}}
\end{equation}
The trace is both over momentum and spinor space, the latter producing a factor of 2. Before bringing forward the heat kernel though, we should proceed carefully, in account of the two distinct infinite sums: the momentum continuum and the discrete sphere modes. We first perform the momentum integration,
\begin{equation}
2\sum_{n}N_n\int\frac{\dd p}{2\pi}\left[\left(\frac{n}{R}\right)^2+p^2\right]^{-s}=\frac{2\,\Gamma\left(s-\frac{1}{2}\right)}{(4\pi)^{\frac{1}{2}}\Gamma(s)}\left(\frac{n^2}{R^2}\right)^{-s+\frac{1}{2}}
\end{equation}
where we set $n=l+1$. This gives
\begin{equation}
\zeta_{\mbox{\tiny$\mathcal{D}^2$}}(s)=\frac{2\,R^{2s-1}}{(4\pi)^{\frac{1}{2}}\Gamma(s)}\int\dd^3x\,\sum_{n}N_n\int_0^\infty\dd\sigma\sigma^{s-\frac{1}{2}-1}e^{-\sigma\frac{n^2}{R^2}}e^{-\sigma\mathcal{E}}
\end{equation}
where we set $n=l+1$ and rescaled $\sigma$ over $R^2$. We also added a convergence term; this is just a lazy, but tinier, alternative to dimensional regularization of momentum. The heat kernel expansion,
\begin{equation}
\int\dd^3x\left(\sum_{n}N_ne^{-\sigma\frac{n^2}{R^2}}\right)\;\;\;\xrightarrow{\sigma\rightarrow0}\;\;\;\frac{1}{(4\pi\sigma)^{\frac{3}{2}}}\sum_{k\geq0}\alpha_k\left(\frac{\sigma}{R^2}\right)^k
\end{equation}
where $k=0,\frac{1}{2},1,\ldots$. Notice that the degeneracy is embedded in the eigenvalue problem, therefore it is part of the kernel-expansion coefficients. Under this expansion,
\begin{equation}
\begin{split}
\zeta_{\mbox{\tiny$\mathcal{D}^2$}}(s)&=\frac{2\,R^{2s-1}}{\Gamma(s)(4\pi)^{2}}\int_0^\infty\dd\sigma\sigma^{s-3}\sum_{k\geq0}\alpha_k\sigma^ke^{-\sigma\mathcal{E}}\\
&=\frac{2\,R^{2s-1}}{\Gamma(s)(4\pi)^{2}}\left\lbrace\alpha_0\Gamma\left(s-2\right)L^{2-s}\;+\;\ldots\right\rbrace
\end{split}
\end{equation}
where we omitted the rest of the terms, since all other coefficients vanish: the half-integer ones represent boundary terms (whereas here we do not have a boundary) and the rest of the integer ones involve potential terms (whereas we consider free fermions). Hence, the sole contribution comes from the leading coefficient,
\begin{equation}
\alpha_{0}=\int\dd^3 x=V_3\;,
\end{equation}
and, using that $\Gamma(s)=(s-1)(s-2)\Gamma(s-2)$,
\begin{equation}
\zeta_{\mbox{\tiny$\mathcal{D}^2$}}(s)=\frac{2\,V_3\,R^{2s-1}\,\mathcal{E}^{2-s}}{(4\pi)^{2}(s-1)(s-2)}\;.
\end{equation}
After we continue to $s=0$, we renormalize the scale $\mathcal{E}=1/R$, letting it also absorb a factor of $\sqrt{4\pi}$ (since the convergence factor, like dimensional regularization, effectively acts like a scale in an extra dimension). Hence, recalling that we have two massless fermions,
\begin{equation}
Z_3\;=\;\exp\left\lbrace-\frac{V_3\,R^{-3}}{(4\pi)^{\frac{3}{2}}}\left(\log\frac{R^2}{\rho^2}+\mbox{finite}\right)\right\rbrace
\end{equation}
where we shifted to a length renormalization scale $\rho=\frac{1}{\mu}$.

\section{Path integral in AdS$_4$}\label{App8}
The universal cover of AdS$_4$ exhibits boundary conditions on its conformal boundary which produce Dirac regular modes, their eigenvalues being \cite{Cotaescu:1998ts}
\begin{equation}
\lambda_n\;=\;\sum N_n\left\lbrace m+R^{-1}\left(n+\frac{3}{2}\right)\right\rbrace,\;\;\;\;\;\;\;n\in\mathbb{N}
\end{equation}
where $m$ is the fermion mass, $R$ is the AdS radius and $N_n=(n+1)(n+2)$ is the degeneracy. As before, we regulate the infinite eigenvalues with a zeta function,
\begin{equation}
\zeta_{\mbox{\tiny$\mathcal{D}^2$}}(s)=\sum_{n}\lambda_{n}^{-2s}=\sum_{n}N_n\left[m^2+R^{-2}\left(n+\frac{3}{2}\right)^2\right]^{-s}\;.
\end{equation}
Introducing a heat kernel, we have
\begin{equation}
\zeta_{\mbox{\tiny$\mathcal{D}^2$}}(s)=\frac{1}{\Gamma(s)}\int\dd^4x\,\sum_{n}N_n\int_0^\infty\dd\sigma\sigma^{s-1}e^{-\sigma\left(m^2+R^{-2}(n+\frac{3}{2})^2\right)}\,.
\end{equation}
Looking forward to the kernel expansion, we should separate the $n$-dependent terms across the eigenvalues. Redefining $n$ to an appropriate $\tilde{n}$, the simplest choice is
\begin{equation}
\zeta_{\mbox{\tiny$\mathcal{D}^2$}}(s)=\frac{1}{\Gamma(s)}\int\dd^4x\,\sum_{\tilde{n}}N_{\tilde{n}}\int_0^\infty\dd\sigma\sigma^{s-1}e^{-\sigma\left(m^2+\lambda_{\tilde{n}}^2+R^{-2}\right)}\,.
\end{equation}
The kernel expands as
\begin{equation}
\int\dd^4x\sum_{\tilde{n}}e^{-\sigma\lambda_{\tilde{n}}^2}\;\;\;\;\xrightarrow{\sigma\rightarrow0}\;\;\;\;\frac{1}{(4\pi\sigma)^2}\sum_{k\geq0}\alpha_k\,\sigma^k\,,
\end{equation}
where $k=0,\frac{1}{2},1,\ldots$. Under this expansion,
\begin{equation}
\begin{split}
\zeta_{\mbox{\tiny$\mathcal{D}^2$}}(s)&=\frac{1}{\Gamma(s)(4\pi)^{2}}\int_0^\infty\dd\sigma\sigma^{s-3}\sum_{k\geq0}\alpha_k\,\sigma^ke^{-\sigma(m^2+R^{-2})}\\
&=\frac{1}{\Gamma(s)(4\pi)^{2}}\left\lbrace\alpha_0\Gamma\left(s-2\right)(m^2+R^{-2})^{2-s}\;+\ldots\right\rbrace
\end{split}
\end{equation}
We did not bother to show the rest of the terms, since it is a free theory and, hence, the sole contribution comes from the leading coefficient, $\alpha_0=\int\dd^4 x=V_A$. Finally, the zeta function,
\begin{equation}
\zeta_{\mbox{\tiny$\mathcal{D}^2$}}(s)=\frac{V_A\,(m^2+R^{-2})^{2-s}}{(4\pi)^{2}(s-1)(s-2)}
\end{equation}
defines the renormalized path integral,
\begin{equation}
Z_A\;=\;\exp\left\lbrace-\frac{1}{4}\frac{V_A\,(m^2+R^{-2})^{2}}{(4\pi)^{2}}\log\left(\frac{\mu^2}{m^2+R^{-2}}\right)\right\rbrace\;.
\end{equation}

\bibliographystyle{apsrev4-1} % Tell bibtex which bibliography style to use
%\bibliography{omega_non_int_kostasNotes.bib} % Tell bibtex which .bib file to use (this one is some example file in TexLive's file tree)

\begin{thebibliography}{99}

%\cite{Filippas:2022sna}
\bibitem{Filippas:2022sna}
K.~Filippas,
``Spin-orbit duality'',
Phys. Rev. D \textbf{108} (2023) no.8, 085029
doi:10.1103/PhysRevD.108.085029
[arXiv:2212.11340 [hep-th]].
%2 citations counted in INSPIRE as of 09 Oct 2024

\bibitem{foot1}
The first is an identity, while the second goes by the name `spin supplementary condition' and may be thought of as setting $S^{\mu\nu}$ as the generators of the little group (see \cite{Filippas:2022sna}).


\bibitem{footN}
The spacetime coordinate $X^\mu$ decomposes w.r.t. $p^\mu$ into $X^\mu=A^\mu+N^\mu$. $A^\mu$ is the aligned part, $A^\mu=(X\cdot p)p^\mu/p^2$, while $N^\mu$ is the normal part, $N^\mu=X^\mu-A^\mu$, $N^\mu p_\mu=0$.


%\cite{Moller}
\bibitem{Moller}
C.~M.~M\o{}ller, ``The Theory of Relativity'', Oxford Univ. Press, (1957);
C.~M.~M\o{}ller, ``Sur la dinamique des syste’mes ayant un moment angulaire interne'', 31 A. I. H. Poincar\`e 11, (1949).

 %\cite{Lorce:2018zpf}
\bibitem{Lorce:2018zpf}
D.~Alba, L.~Lusanna and M.~Pauri,
``Centers of mass and rotational kinematics for the relativistic N body problem in the rest frame instant form'',
J. Math. Phys. \textbf{43} (2002), 1677-1727,
doi:10.1063/1.1435424
[arXiv:hep-th/0102087 [hep-th]];
%45 citations counted in INSPIRE as of 08 Dec 2022
C.~Lorc\'e,
``The relativistic center of mass in field theory with spin'',
Eur. Phys. J. C \textbf{78} (2018) no.9, 785,
doi:10.1140/epjc/s10052-018-6249-3
[arXiv:1805.05284 [hep-ph]].
%24 citations counted in INSPIRE as of 08 Dec 2022


%\cite{Pryce:1948}
\bibitem{Pryce:1948}
M.~H.~L.~Pryce,
``The mass-centre in the restricted theory of relativity and its connexion with the quantum theory of elementary particles'',
P. R. Soc. Lond. A19562–81, Vol 195, Issue 1040.


%\cite{Casalbuoni:1976tz}
\bibitem{Casalbuoni:1976tz}
R.~Casalbuoni,
``The Classical Mechanics for Bose-Fermi Systems'',
Nuovo Cim. A \textbf{33} (1976), 389
doi:10.1007/BF02729860.
%454 citations counted in INSPIRE as of 11 Mar 2024

%\cite{Brink:1981nb}
\bibitem{Brink:1981nb}
L.~Brink and J.~H.~Schwarz,
``Quantum Superspace'',
Phys. Lett. B \textbf{100} (1981), 310-312
doi:10.1016/0370-2693(81)90093-9.
%439 citations counted in INSPIRE as of 11 Mar 2024

%\cite{Lukierski}
\bibitem{Lukierski}
J.~Lukierski, H.~Ruegg, A.~Nowicki and V.~N.~Tolstoi,
``Q deformation of Poincare algebra'',
Phys. Lett. B \textbf{264} (1991), 331-338
doi:10.1016/0370-2693(91)90358-W;
%725 citations counted in INSPIRE as of 09 Dec 2022
J.~Lukierski, A.~Nowicki and H.~Ruegg,
``New quantum Poincare algebra and k deformed field theory'',
Phys. Lett. B \textbf{293} (1992), 344-352
doi:10.1016/0370-2693(92)90894-A.
%514 citations counted in INSPIRE as of 09 Dec 2022

\bibitem{footSpinNC}
In this expression for the dual noncommutative algebra, the timelike part of $\hat{X}^\mu$ cancels out completely and only the spacelike $\hat{W}^\mu$ remains \cite{Filippas:2022sna}. In this form, the so-called spin-noncommutative algebra is realized \cite{Falomir:2009cq}.


%\cite{Falomir:2009cq}
\bibitem{Falomir:2009cq}
H.~Falomir, J.~Gamboa, J.~Lopez-Sarrion, F.~Mendez and P.~A.~G.~Pisani,
%``Magnetic-Dipole Spin Effects in Noncommutative Quantum Mechanics'',
Phys. Lett. B \textbf{680} (2009), 384-386
doi:10.1016/j.physletb.2009.09.007
[arXiv:0905.0157 [hep-th]];
A.~Das, H.~Falomir, M.~Nieto, J.~Gamboa and F.~Mendez,
%``Aharonov-Bohm effect in a Class of Noncommutative Theories,''
Phys. Rev. D \textbf{84} (2011), 045002
doi:10.1103/PhysRevD.84.045002
[arXiv:1105.1800 [hep-th]];
%29 citations counted in INSPIRE as of 08 Dec 2022
%\cite{Deriglazov:2013zaa}
D.~Karabali and V.~P.~Nair,
%``Relativistic Particle and Relativistic Fluids: Magnetic Moment and Spin-Orbit Interactions,''
Phys. Rev. D \textbf{90} (2014) no.10, 105018
doi:10.1103/PhysRevD.90.105018
[arXiv:1406.1551 [hep-th]];
A.~A.~Deriglazov and W.~Guzm\'an Ram\'\i{}rez,
%``Recent progress on the description of relativistic spin: vector model of spinning particle and rotating body with gravimagnetic moment in General Relativity,''
Adv. Math. Phys. \textbf{2017} (2017), 7397159
doi:10.1155/2017/7397159
[arXiv:1710.07135 [gr-qc]];
R.~Jackiw and V.~P.~Nair,
%``Anyon spin and the exotic central extension of the planar Galilei group'',
Phys. Lett. B \textbf{480} (2000), 237-238
doi:10.1016/S0370-2693(00)00379-8
[arXiv:hep-th/0003130 [hep-th]];
M.~Gomes, V.~G.~Kupriyanov and A.~J.~da Silva,
%``Noncommutativity due to spin'',
Phys. Rev. D \textbf{81} (2010), 085024
doi:10.1103/PhysRevD.81.085024
[arXiv:1002.4173 [hep-th]];
A.~F.~Ferrari, M.~Gomes, V.~G.~Kupriyanov and C.~A.~Stechhahn,
%``Dynamics of a Dirac Fermion in the presence of spin noncommutativity'',
Phys. Lett. B \textbf{718} (2013), 1475-1480
doi:10.1016/j.physletb.2012.12.010
[arXiv:1207.0412 [hep-th]];
J.~Lukierski and M.~Woronowicz,
%``Noncommutative Space-time from Quantized Twistors'',
doi:10.1142/9789814590112\_0025
[arXiv:1311.7498 [hep-th]].
%2 citations counted in INSPIRE as of 08 Dec 2022





%\cite{Newton:1949cq}
\bibitem{Newton:1949cq}
T.~D.~Newton and E.~P.~Wigner,
``Localized States for Elementary Systems'',
Rev. Mod. Phys. \textbf{21} (1949), 400-406
doi:10.1103/RevModPhys.21.400.
%790 citations counted in INSPIRE as of 09 Dec 2022

%\cite{Madore:1991bw}
\bibitem{Madore:1991bw}
J.~Madore,
``The Fuzzy sphere'',
Class. Quant. Grav. \textbf{9} (1992), 69-88
doi:10.1088/0264-9381/9/1/008.
%646 citations counted in INSPIRE as of 09 Dec 2022




%\cite{Bose}
\bibitem{Bose}
S.~Bose and D.~Home,
``Duality in Entanglement Enabling a Test of Quantum Indistinguishability Unaffected by Interactions'',
Phys. Rev. Lett. \textbf{110} (2013), 140404
doi:10.1103/physrevlett.110.140404
[arXiv:1304.1435 [quant-ph]];

%\cite{Moreva}
\bibitem{Moreva}
E.~Moreva, G.~Brida and M.~Gramegna, S.~Bose, D.~Home and M.~Genovese,
``Bell measurements as a witness of a dualism in entanglement'',
Phys. Rev. A \textbf{91} (2015), 062117
doi:10.1103/PhysRevA.91.062117
[arXiv:1503.06153 [quant-ph]];
J-J.~Ma et al, 2014 New J. Phys. 16 083011
[arXiv:1408.5190 [quant-ph]];

%\cite{Karczewski}
\bibitem{Karczewski}
M.~Karczewski and P.~Kurzy\'nski,
``How to observe duality in entanglement of two distinguishable particles'',
Phys. Rev. A \textbf{94} (2016), 032124
doi:10.1103/PhysRevA.94.032124
[arXiv:1608.01525 [quant-ph]];

%\cite{Bhatti}
\bibitem{Bhatti}
D.~Bhatti, J.~von~Zanthier and G.~S.~Agarwal,
``Entanglement of polarization and orbital angular momentum'',
Phys. Rev. A \textbf{91} (2015), 062303
doi:10.1103/PhysRevA.91.062303
[arXiv:1502.01906 [quant-ph]];
S.~Lee, C.~Lee, P.~Kurzy\'nski, D.~Kaszlikowski and J,~Kim,
``Duality in entanglement of macroscopic states of light'',
Phys. Rev. A \textbf{94} (2016), 022314
doi:10.1103/PhysRevA.94.022314
[arXiv:1606.01613 [quant-ph]].




\bibitem{footBMS}
Note that, in the literature, super-translations are given on the asymptotically-flat spacetime. Here, they are found on the dual space, i.e. null infinity, itself. That is, while time super-translation is the same (since the time coordinate is invariant under the duality), spatial ones differ and their exact form is found via the Hopf map.



%\cite{Bondi:1962px}
\bibitem{Bondi:1962px}
H.~Bondi, M.~G.~J.~van der Burg and A.~W.~K.~Metzner,
``Gravitational waves in general relativity. 7. Waves from axisymmetric isolated systems'',
Proc. Roy. Soc. Lond. A \textbf{269} (1962), 21-52
doi:10.1098/rspa.1962.0161;
%1899 citations counted in INSPIRE as of 25 Oct 2024
R.~K.~Sachs,
``Gravitational waves in general relativity. 8. Waves in asymptotically flat space-times'',
Proc. Roy. Soc. Lond. A \textbf{270} (1962), 103-126
doi:10.1098/rspa.1962.0206.
%1429 citations counted in INSPIRE as of 25 Oct 2024


%\cite{deBoer:2003vf}
\bibitem{deBoer:2003vf}
J.~de Boer and S.~N.~Solodukhin,
``A Holographic reduction of Minkowski space-time'',
Nucl. Phys. B \textbf{665} (2003), 545-593
doi:10.1016/S0550-3213(03)00494-2
[arXiv:hep-th/0303006 [hep-th]].
%291 citations counted in INSPIRE as of 25 Oct 2024

%\cite{footHopfIns}
\bibitem{footHopfIns}
A similar quantum-mechanical example that involves charges in a magnetic field and the Hopf map, and should thus apply here, are the magnetic `Hopf insulators' \cite{Moore2008}.



%\cite{Moore2008}
\bibitem{Moore2008}
J.~E.~Moore, Y.~Ran, X.~-G.~Wen,
``Topological surface states in three-dimensional magnetic insulators'',
Phys. Rev. Lett. \textbf{101} (2008),
doi:10.1103/PhysRevLett.101.186805;
D.~-L.~Deng, S.~-T.~Wang, C.~Shen, L.~-M. Duan,
``Hopf insulators and their topologically protected surface states'',
Phys. Rev. B \textbf{88} (2013),
doi:10.1103/PhysRevB.88.201105


\bibitem{foot2}
Besides realizing the Hopf fibration, this dipole/monopole correspondence could be expected, since the duality is an interchange of a divergent-less (spin) and a curl-free (orbital) part of a Hodge decomposition.



%\cite{Minami:1979wn}
\bibitem{Minami:1979wn}
M.~Minami,
``Dirac's Monopole and the Hopf Map'',
Prog. Theor. Phys. \textbf{62} (1979), 1128
doi:10.1143/PTP.62.1128;
K.~Hasebe,
``Hopf Maps, Lowest Landau Level, and Fuzzy Spheres'',
SIGMA \textbf{6} (2010), 071
doi:10.3842/SIGMA.2010.071
[arXiv:1009.1192 [hep-th]].
%38 citations counted in INSPIRE as of 12 May 2024

%\cite{Haldane:1983xm}
\bibitem{Haldane:1983xm}
F.~D.~M.~Haldane,
``Fractional quantization of the Hall effect: A Hierarchy of incompressible quantum fluid states'',
Phys. Rev. Lett. \textbf{51} (1983), 605-608
doi:10.1103/PhysRevLett.51.605
%933 citations counted in INSPIRE as of 23 Oct 2024





\bibitem{foot3}
Depending on the gauge, the Hamiltonian involves analogous oscillator and angular-momentum coordinates.

\bibitem{foot4}
The Lorentzian map (\ref{MAP}) is equal to $L_{\mu\nu}\mapsto\star S_{\mu\nu}$, i.e. $L_i=\frac{1}{2}\epsilon_{ijk}L^{jk}\mapsto S_{0i}$. The Euclidean decomposition $\mbox{SO}(4)\cong\mbox{SU}(2)\times\mbox{SU}(2)$ produces a subalgebra on the generators $\widetilde{S}_i=S_{4i}+\frac{1}{2}\epsilon_{ijk}S^{jk}$, which implies $S_{4i}=\widetilde{S}_i-\frac{1}{2}\epsilon_{ijk}S^{jk}:=S_i$. In this context, where we just aim for a qualitative view, we consider both spin vectors coming from mapping $X^i$ or $L^i$ to be the same.


%\cite{Kopec}
\bibitem{Kopec}
Kope\ifmmode \acute{c}\else \'{c}\fi{}, T. K. and Usadel, K. D.,
``Quantum Spin Glass on the Bethe Lattice'',
Phys. Rev. Lett. \textbf{78} (1997), 1988-1991
doi:10.1103/PhysRevLett.78.1988.


%\cite{Nareddy}
\bibitem{Nareddy}
V. R. Nareddy and J. Machta,
``Kinetic Ising Models with Self-interaction: Sequential and Parallel Updating'',
Phys. Rev. E \textbf{101} (2020), 012122
doi:10.1103/PhysRevE.101.012122.


%\cite{Hiley}
\bibitem{Hiley}
B. J. Hiley and G. S. Joyce,
``The Ising model with long-range interactions'',
1965 Proc. Phys. Soc. 85 493
doi: 10.1088/0370-1328/85/3/310.


%\cite{Paulos:2015jfa}
\bibitem{Paulos:2015jfa}
M.~F.~Paulos, S.~Rychkov, B.~C.~van Rees and B.~Zan,
``Conformal Invariance in the Long-Range Ising Model'',
Nucl. Phys. B \textbf{902} (2016), 246-291
doi:10.1016/j.nuclphysb.2015.10.018
[arXiv:1509.00008 [hep-th]].
%98 citations counted in INSPIRE as of 03 May 2024


%\cite{Zhu:2022gjc}
\bibitem{Zhu:2022gjc}
W.~Zhu, C.~Han, E.~Huffman, J.~S.~Hofmann and Y.~C.~He,
``Uncovering Conformal Symmetry in the 3D Ising Transition: State-Operator Correspondence from a Quantum Fuzzy Sphere Regularization'',
Phys. Rev. X \textbf{13} (2023) no.2, 021009
doi:10.1103/PhysRevX.13.021009
[arXiv:2210.13482 [cond-mat.stat-mech]].
%30 citations counted in INSPIRE as of 11 Apr 2024



\bibitem{foot5}
Since the dual space is really $[\mathbb{S}^2\times\mathbb{R}]$, this spin connection would actually be a conformal connection, Weyl-transforming across the conformal class of metrics.


\bibitem{foot6}
The factor $1/2$ is because the generator acts on $\Psi^\mu=\gamma^\mu\psi/2m$. $\mathfrak{L}^{\mu\nu}$ is already set appropriately.

%\cite{Borokhov:2002ib}
\bibitem{Borokhov:2002ib}
V.~Borokhov, A.~Kapustin and X.~k.~Wu,
``Topological disorder operators in three-dimensional conformal field theory'',
JHEP \textbf{11} (2002), 049
doi:10.1088/1126-6708/2002/11/049
[arXiv:hep-th/0206054 [hep-th]].
%274 citations counted in INSPIRE as of 08 Dec 2024


\bibitem{foot9}
This is the classical radius $R$ of the cylinder, i.e. (\ref{MollerRadius}), where $R=1/m$ reflects fermion number equal to one.

\bibitem{foot10}
At a different point of the conformal class, e.g. Weyl-transforming to $\mathbb{R}^3$, a conformal boundary emerges which should produce the above result, the contribution coming solely from the conformal anomaly.


\bibitem{footVasiliev}
Dual massive fermions in AdS$_4$ is obviously an example of bottom-up holography, outside the standard AdS/CFT paradigm. The vacuum may generalize to $N_f$ fermions (and $2N_f$ at null infinity). In this case, for $N_f\rightarrow\infty$, it has been shown \cite{Sezgin:2003pt} that the free fermionic CFT is dual to Vasiliev's higher-spin B-type theory in AdS$_4$ \cite{Vasiliev:2003ev}. Even if the full theory is captured by higher-spin currents, this does not prevent us, in any way, from looking at massive fermions in AdS$_4$ as a dual free vacuum. Besides, our calculation is valid for any $N_f$.

%\cite{Sezgin:2003pt}
\bibitem{Sezgin:2003pt}
E.~Sezgin and P.~Sundell,
``Holography in 4D (super) higher spin theories and a test via cubic scalar couplings'',
JHEP \textbf{07} (2005), 044
doi:10.1088/1126-6708/2005/07/044
[arXiv:hep-th/0305040 [hep-th]];
%332 citations counted in INSPIRE as of 27 Oct 2024
S.~Giombi and X.~Yin,
``The Higher Spin/Vector Model Duality'',
J. Phys. A \textbf{46} (2013), 214003
doi:10.1088/1751-8113/46/21/214003
[arXiv:1208.4036 [hep-th]];
%251 citations counted in INSPIRE as of 27 Oct 2024
S.~Giombi, V.~Kirilin and E.~Skvortsov,
``Notes on Spinning Operators in Fermionic CFT'',
JHEP \textbf{05} (2017), 041
doi:10.1007/JHEP05(2017)041
[arXiv:1701.06997 [hep-th]].
%59 citations counted in INSPIRE as of 27 Oct 2024

%\cite{Vasiliev:2003ev}
\bibitem{Vasiliev:2003ev}
M.~A.~Vasiliev,
``Nonlinear equations for symmetric massless higher spin fields in (A)dS(d)'',
Phys. Lett. B \textbf{567} (2003), 139-151
doi:10.1016/S0370-2693(03)00872-4
[arXiv:hep-th/0304049 [hep-th]].
%611 citations counted in INSPIRE as of 27 Oct 2024


%\cite{Camporesi:1995fb}
\bibitem{Camporesi:1995fb}
R.~Camporesi and A.~Higuchi,
``On the Eigen functions of the Dirac operator on spheres and real hyperbolic spaces'',
J. Geom. Phys. \textbf{20} (1996), 1-18
doi:10.1016/0393-0440(95)00042-9
[arXiv:gr-qc/9505009 [gr-qc]].
%247 citations counted in INSPIRE as of 23 Oct 2024



%\cite{Arcioni:2003xx}
\bibitem{Arcioni:2003xx}
G.~Arcioni and C.~Dappiaggi,
%``Exploring the holographic principle in asymptotically flat space-times via the BMS group,''
Nucl. Phys. B \textbf{674} (2003), 553-592
doi:10.1016/j.nuclphysb.2003.09.051
[arXiv:hep-th/0306142 [hep-th]];
A.~Bagchi,
%``Correspondence between Asymptotically Flat Spacetimes and Nonrelativistic Conformal Field Theories,''
Phys. Rev. Lett. \textbf{105} (2010), 171601
doi:10.1103/PhysRevLett.105.171601
[arXiv:1006.3354 [hep-th]];
A.~Bagchi and R.~Fareghbal,
%``BMS/GCA Redux: Towards Flatspace Holography from Non-Relativistic Symmetries,''
JHEP \textbf{10} (2012), 092
doi:10.1007/JHEP10(2012)092
[arXiv:1203.5795 [hep-th]];
R.~B.~Mann and D.~Marolf,
%``Holographic renormalization of asymptotically flat spacetimes,''
Class. Quant. Grav. \textbf{23} (2006), 2927-2950
doi:10.1088/0264-9381/23/9/010
[arXiv:hep-th/0511096 [hep-th]].
%150 citations counted in INSPIRE as of 29 Oct 2024



%\cite{Cheung:2016iub}
\bibitem{Cheung:2016iub}
C.~Cheung, A.~de la Fuente and R.~Sundrum,
%``4D scattering amplitudes and asymptotic symmetries from 2D CFT,''
JHEP \textbf{01} (2017), 112
doi:10.1007/JHEP01(2017)112
[arXiv:1609.00732 [hep-th]];
S.~Pasterski, S.~H.~Shao and A.~Strominger,
%``Flat Space Amplitudes and Conformal Symmetry of the Celestial Sphere,''
Phys. Rev. D \textbf{96} (2017) no.6, 065026
doi:10.1103/PhysRevD.96.065026
[arXiv:1701.00049 [hep-th]];
A.~Ball, E.~Himwich, S.~A.~Narayanan, S.~Pasterski and A.~Strominger,
%``Uplifting AdS$_{3}$/CFT$_{2}$ to flat space holography,''
JHEP \textbf{08} (2019), 168
doi:10.1007/JHEP08(2019)168
[arXiv:1905.09809 [hep-th]];
L.~Donnay, A.~Puhm and A.~Strominger,
%``Conformally Soft Photons and Gravitons,''
JHEP \textbf{01} (2019), 184
doi:10.1007/JHEP01(2019)184
[arXiv:1810.05219 [hep-th]];
W.~Fan, A.~Fotopoulos and T.~R.~Taylor,
%``Soft Limits of Yang-Mills Amplitudes and Conformal Correlators,''
JHEP \textbf{05} (2019), 121
doi:10.1007/JHEP05(2019)121
[arXiv:1903.01676 [hep-th]];
A.~Fotopoulos and T.~R.~Taylor,
%``Primary Fields in Celestial CFT,''
JHEP \textbf{10} (2019), 167
doi:10.1007/JHEP10(2019)167
[arXiv:1906.10149 [hep-th]];
L.~Donnay, S.~Pasterski and A.~Puhm,
%``Asymptotic Symmetries and Celestial CFT,''
JHEP \textbf{09} (2020), 176
doi:10.1007/JHEP09(2020)176
[arXiv:2005.08990 [hep-th]];
A.~Guevara, E.~Himwich, M.~Pate and A.~Strominger,
%``Holographic symmetry algebras for gauge theory and gravity,''
JHEP \textbf{11} (2021), 152
doi:10.1007/JHEP11(2021)152
[arXiv:2103.03961 [hep-th]];
J.~Mago, L.~Ren, A.~Y.~Srikant and A.~Volovich,
%``Deformed $w_{1+\infty}$ Algebras in the Celestial CFT,''
SIGMA \textbf{19} (2023), 044
doi:10.3842/SIGMA.2023.044
[arXiv:2111.11356 [hep-th]];
K.~Nguyen,
%``Schwarzian transformations at null infinity,''
PoS \textbf{CORFU2021} (2022), 133
doi:10.22323/1.406.0133
[arXiv:2201.09640 [hep-th]];
L.~Donnay, A.~Fiorucci, Y.~Herfray and R.~Ruzziconi,
%``Carrollian Perspective on Celestial Holography,''
Phys. Rev. Lett. \textbf{129} (2022) no.7, 071602
doi:10.1103/PhysRevLett.129.071602
[arXiv:2202.04702 [hep-th]].
%160 citations counted in INSPIRE as of 29 Oct 2024


%\cite{Dappiaggi:2005ci}
\bibitem{Dappiaggi:2005ci}
C.~Dappiaggi, V.~Moretti and N.~Pinamonti,
%``Rigorous steps towards holography in asymptotically flat spacetimes,''
Rev. Math. Phys. \textbf{18} (2006), 349-416
doi:10.1142/S0129055X0600270X
[arXiv:gr-qc/0506069 [gr-qc]];
C.~Duval, G.~W.~Gibbons and P.~A.~Horvathy,
%``Conformal Carroll groups and BMS symmetry,''
Class. Quant. Grav. \textbf{31} (2014), 092001
doi:10.1088/0264-9381/31/9/092001
[arXiv:1402.5894 [gr-qc]];
A.~Bagchi, R.~Basu, A.~Kakkar and A.~Mehra,
%``Flat Holography: Aspects of the dual field theory,''
JHEP \textbf{12} (2016), 147
doi:10.1007/JHEP12(2016)147
[arXiv:1609.06203 [hep-th]];
A.~Bagchi, A.~Mehra and P.~Nandi,
%``Field Theories with Conformal Carrollian Symmetry,''
JHEP \textbf{05} (2019), 108
doi:10.1007/JHEP05(2019)108
[arXiv:1901.10147 [hep-th]];
A.~Laddha, S.~G.~Prabhu, S.~Raju and P.~Shrivastava,
%``The Holographic Nature of Null Infinity,''
SciPost Phys. \textbf{10} (2021) no.2, 041
doi:10.21468/SciPostPhys.10.2.041
[arXiv:2002.02448 [hep-th]];
B.~Chen, R.~Liu and Y.~f.~Zheng,
%``On higher-dimensional Carrollian and Galilean conformal field theories,''
SciPost Phys. \textbf{14} (2023) no.5, 088
doi:10.21468/SciPostPhys.14.5.088
[arXiv:2112.10514 [hep-th]];
A.~Bagchi, D.~Grumiller and P.~Nandi,
%``Carrollian superconformal theories and super BMS,''
JHEP \textbf{05} (2022), 044
doi:10.1007/JHEP05(2022)044
[arXiv:2202.01172 [hep-th]].
%30 citations counted in INSPIRE as of 29 Oct 2024



\bibitem{MinimalCoupling}
In fact, it seems that $\lambda=g$ for a minimal coupling between the scalar and gauge fields.


%\cite{Maldacena:1997re}
\bibitem{Maldacena:1997re}
J.~M.~Maldacena,
``The Large N limit of superconformal field theories and supergravity'',
Adv. Theor. Math. Phys. \textbf{2} (1998), 231-252
doi:10.4310/ATMP.1998.v2.n2.a1
[arXiv:hep-th/9711200 [hep-th]];
%20158 citations counted in INSPIRE as of 30 Oct 2024
E.~Witten,
``Anti-de Sitter space and holography'',
Adv. Theor. Math. Phys. \textbf{2} (1998), 253-291
doi:10.4310/ATMP.1998.v2.n2.a2
[arXiv:hep-th/9802150 [hep-th]].
%12975 citations counted in INSPIRE as of 07 Dec 2024

%\cite{Lust:2019zwm}
\bibitem{Lust:2019zwm}
D.~L\"ust, E.~Palti and C.~Vafa,
%``AdS and the Swampland,''
Phys. Lett. B \textbf{797} (2019), 134867
doi:10.1016/j.physletb.2019.134867
[arXiv:1906.05225 [hep-th]];
Y.~Li, E.~Palti and N.~Petri,
%``Towards AdS distances in string theory,''
JHEP \textbf{08} (2023), 210
doi:10.1007/JHEP08(2023)210
[arXiv:2306.02026 [hep-th]];
A.~Mohseni, M.~Montero, C.~Vafa and I.~Valenzuela,
%``On Measuring Distances in the Quantum Gravity Landscape,''
[arXiv:2407.02705 [hep-th]];
T.~Rudelius,
%``Dimensional reduction and (Anti) de Sitter bounds,''
JHEP \textbf{08} (2021), 041
doi:10.1007/JHEP08(2021)041
[arXiv:2101.11617 [hep-th]];
D.~van de Heisteeg, C.~Vafa and M.~Wiesner,
%``Bounds on Species Scale and the Distance Conjecture,''
Fortsch. Phys. \textbf{71} (2023) no.10-11, 2300143
doi:10.1002/prop.202300143
[arXiv:2303.13580 [hep-th]].
%45 citations counted in INSPIRE as of 01 Nov 2024



%\cite{Gopakumar:1998ki}
\bibitem{Gopakumar:1998ki}
R.~Gopakumar and C.~Vafa,
%``On the gauge theory / geometry correspondence,''
Adv. Theor. Math. Phys. \textbf{3} (1999), 1415-1443
doi:10.4310/ATMP.1999.v3.n5.a5
[arXiv:hep-th/9811131 [hep-th]];
%713 citations counted in INSPIRE as of 22 Nov 2024
P.~Haggi-Mani and B.~Sundborg,
%``Free large N supersymmetric Yang-Mills theory as a string theory,''
JHEP \textbf{04} (2000), 031
doi:10.1088/1126-6708/2000/04/031
[arXiv:hep-th/0002189 [hep-th]];
%116 citations counted in INSPIRE as of 22 Nov 2024
D.~E.~Berenstein, J.~M.~Maldacena and H.~S.~Nastase,
%``Strings in flat space and pp waves from N=4 superYang-Mills,''
JHEP \textbf{04} (2002), 013
doi:10.1088/1126-6708/2002/04/013
[arXiv:hep-th/0202021 [hep-th]];
%2035 citations counted in INSPIRE as of 22 Nov 2024
A.~Karch,
%``Light cone quantization of string theory duals of free field theories,''
[arXiv:hep-th/0212041 [hep-th]];
%22 citations counted in INSPIRE as of 22 Nov 2024
A.~Dhar, G.~Mandal and S.~R.~Wadia,
%``String bits in small radius AdS and weakly coupled N=4 superYang-Mills theory. 1.,''
[arXiv:hep-th/0304062 [hep-th]];
%39 citations counted in INSPIRE as of 22 Nov 2024
R.~Gopakumar,
%``From free fields to AdS,''
Phys. Rev. D \textbf{70} (2004), 025009
doi:10.1103/PhysRevD.70.025009
[arXiv:hep-th/0308184 [hep-th]];
%153 citations counted in INSPIRE as of 22 Nov 2024
E.~Witten,
%``Perturbative gauge theory as a string theory in twistor space,''
Commun. Math. Phys. \textbf{252} (2004), 189-258
doi:10.1007/s00220-004-1187-3
[arXiv:hep-th/0312171 [hep-th]];
%1346 citations counted in INSPIRE as of 22 Nov 2024
N.~Berkovits,
%``A New Limit of the AdS(5) x S**5 Sigma Model,''
JHEP \textbf{08} (2007), 011
doi:10.1088/1126-6708/2007/08/011
[arXiv:hep-th/0703282 [hep-th]];
%41 citations counted in INSPIRE as of 22 Nov 2024
N.~Berkovits and C.~Vafa,
%``Towards a Worldsheet Derivation of the Maldacena Conjecture,''
JHEP \textbf{03} (2008), 031
doi:10.1088/1126-6708/2008/03/031
[arXiv:0711.1799 [hep-th]];
%54 citations counted in INSPIRE as of 22 Nov 2024
N.~Berkovits,
%``Perturbative Super-Yang-Mills from the Topological AdS(5) x S**5 Sigma Model,''
JHEP \textbf{09} (2008), 088
doi:10.1088/1126-6708/2008/09/088
[arXiv:0806.1960 [hep-th]];
%35 citations counted in INSPIRE as of 22 Nov 2024
N.~Berkovits,
%``Sketching a Proof of the Maldacena Conjecture at Small Radius,''
JHEP \textbf{06} (2019), 111
doi:10.1007/JHEP06(2019)111
[arXiv:1903.08264 [hep-th]].
%24 citations counted in INSPIRE as of 22 Nov 2024


%\cite{footQin}
\bibitem{footQin}
There is a natural correspondence with the space of spatial four-momenta, which is also $\mathbb{R}^3\setminus\lbrace0\rbrace$, since $p^\mu=0$ is not propagating and is removed. See \cite{Palmerduca:2023ctx} for an application on massless fields.

%\cite{Palmerduca:2023ctx}
\bibitem{Palmerduca:2023ctx}
E.~Palmerduca and H.~Qin,
``Photon topology'',
Phys. Rev. D \textbf{109} (2024) no.8, 085005
doi:10.1103/PhysRevD.109.085005
[arXiv:2308.11147 [math-ph]];
%5 citations counted in INSPIRE as of 16 Dec 2024
E.~Palmerduca and H.~Qin,
``Graviton topology'',
JHEP \textbf{11} (2024), 150
doi:10.1007/JHEP11(2024)150
[arXiv:2404.11696 [math-ph]].
%2 citations counted in INSPIRE as of 16 Dec 2024



%\cite{Angelone:2022tet}
\bibitem{Angelone:2022tet}
G.~Angelone, E.~Ercolessi, P.~Facchi, D.~Lonigro, R.~Maggi, G.~Marmo, S.~Pascazio and F.~V.~Pepe,
``Dimensional reduction of the Dirac theory'',
J. Phys. A \textbf{56} (2023) no.6, 065201
doi:10.1088/1751-8121/acb869
[arXiv:2211.08581 [quant-ph]];
%4 citations counted in INSPIRE as of 23 Oct 2024
D.~Lonigro, R.~Maggi, G.~Angelone, E.~Ercolessi, P.~Facchi, G.~Marmo, S.~Pascazio and F.~V.~Pepe,
``Dimensional reduction of the Dirac equation in arbitrary spatial dimensions'',
Eur. Phys. J. Plus \textbf{138} (2023) no.4, 324
doi:10.1140/epjp/s13360-023-03919-0
[arXiv:2212.11965 [quant-ph]].
%0 citations counted in INSPIRE as of 23 Oct 2024



%\cite{Bar}
\bibitem{Bar}
C.~Bar,
``The Dirac operator on space forms of positive curvature'',
J. Math. Soc. Japan 48(1): 69-83 (January, 1996). DOI: 10.2969/jmsj/04810069


%\cite{Cotaescu:1998ts}
\bibitem{Cotaescu:1998ts}
I.~I.~Cotaescu,
``The Dirac particle on central backgrounds and the anti-de Sitter oscillator'',
Mod. Phys. Lett. A \textbf{13} (1998), 2923-2936
doi:10.1142/S0217732398003107
[arXiv:gr-qc/9803042 [gr-qc]];
%24 citations counted in INSPIRE as of 24 Oct 2024
I.~I.~Cotaescu,
``Dirac fermions in de Sitter and anti-de Sitter backgrounds'',
Rom. J. Phys. \textbf{52} (2007), 895-940
[arXiv:gr-qc/0701118 [gr-qc]].
%19 citations counted in INSPIRE as of 24 Oct 2024



\end{thebibliography}

\end{document}